\documentclass[english]{article}
\usepackage{lmodern}
\usepackage[T1]{fontenc}
\usepackage[latin9]{inputenc}
\usepackage{geometry}
\usepackage{pdfpages}
\usepackage{amsmath}
\usepackage{amssymb}
\usepackage{setspace}
\usepackage[authoryear]{natbib}
\usepackage{graphicx}
\usepackage[justification=centering]{subfig}
\PassOptionsToPackage{normalem}{ulem}
\usepackage{ulem}
\makeatletter

\AtBeginDocument{
\usepackage{lineno}
\addtolength{\abovedisplayskip}{-1.0ex}
\addtolength{\abovedisplayshortskip}{-1.0ex}
\addtolength{\belowdisplayskip}{-1.0ex}
\addtolength{\belowdisplayshortskip}{-1.0ex}
}

\usepackage{changepage}
\usepackage[hidelinks]{hyperref}

\makeatother

\usepackage{babel}
\begin{document}
\begin{singlespace}

\title{\vspace{1.2cm}Estimating Treatment Effects in Mover Designs\thanks{I thank Sarah Abraham, Alberto Abadie, Isaiah Andrews, Josh Angrist, Amy Finkelstein, Matt Gentzkow, Paul Goldsmith-Pinkham, Sida Peng, Nathan Hendren, Liyang Sun, Chris Walters, and Heidi Williams for valuable feedback. This draft is a work in progress and should not be cited without prior notification.}
\bigskip{}
}
\end{singlespace}

\author{Peter Hull%
\thanks{University of Chicago and Microsoft Research. Email: \href{mailto:hull@uchicago.edu}{\nolinkurl{hull@uchicago.edu}}; website: \url{http://peterhull.net}
}}
\date{April 2018}
\maketitle
\begin{abstract}
\begin{spacing}{1.1}
\begin{adjustwidth*}{0.4cm}{0.4cm} {\normalsize Researchers increasingly leverage movement across multiple treatments to estimate causal effects. While these ``mover regressions'' are often motivated by a linear constant-effects model, it is not clear what they capture under weaker quasi-experimental assumptions. I show that binary treatment mover regressions recover a convex average of four difference-in-difference comparisons and are thus causally interpretable under a standard parallel trends assumption. Estimates from multiple-treatment models, however, need not be causal without stronger restrictions on the heterogeneity of treatment effects and time-varying shocks. I propose a class of two-step estimators to isolate and combine the large set of difference-in-difference quasi-experiments generated by a mover design, identifying mover average treatment effects under conditional-on-covariate parallel trends and effect homogeneity restrictions. I characterize the efficient estimators in this class and derive specification tests based on the model's overidentifying restrictions. Future drafts will apply the theory to the Finkelstein et al. (2016) movers design, analyzing the causal effects of geography on healthcare utilization.}\end{adjustwidth*} \thispagestyle{empty}\vfill{}
\end{spacing}
\pagebreak{}

\end{abstract}

\section{Introduction}
\setcounter{page}{1}

The rise of rich longitudinal data has broadened the scope for causal inference in economics. Rather than estimating a single cross-sectional treatment effect, researchers increasingly exploit variation in an individual's treatment choices over time -- such as the firm they work for, the city they live in, the teacher they learn from, or the doctor they are treated by -- in order to estimate a large number of causal effects simultaneously.\footnote{Recent examples include \cite{bdg12}, \cite{chk_2013}, \cite{jackson13}, \cite{cfr:14a}, \cite{fui2015}, \cite{fgw16}, \cite{sacarnyjmp}, \cite{fghw17}, \cite{molitor17}, \cite{alcottetal17}, and \cite{chetty/hendren:15}.} Often these effects are obtained from a linear two-way fixed effects regression, motivated by a static, additive, and constant-effect model (e.g. \cite{akm99}). It is usually not clear, however, what these sorts of ``mover regressions'' capture under misspecifications of the model, including heterogeneous treatment effects, outcome persistence, and time-varying shocks.

This paper explores the causal content of mover designs in a treatment effects framework, relaxing the canonical regression model with familiar quasi-experimental restrictions. In the simplest case of a binary treatment, two time periods, and no additional controls, I show that a mover regression identifies a weighted average of four difference-in-difference comparisons, and is therefore causally interpretable under a restriction on outcome persistance and a standard parallel trends assumption. This result links mover analyses to simpler quasi-experimental designs, but does not easily extend to settings with multiple unordered treatments. I show that, in general, mover regressions need not identify weighted averages of heterogeneous causal effects under parallel trends alone; rather they require additional restrictions on potential outcome heterogeneity across both treatments and time. 

Motivated by these results, I develop a class of two-step \emph{mover average treatment effect} (MATE) estimators for quasi-experimental mover designs. The estimators can be thought to extend the semiparametric difference-in-difference approach of \cite{abadie_05} to settings where individuals move both into and out of multiple treatments over time. Identification follows from conditional-on-covariate restrictions on trends and treatment effect heterogeneity, which are satisfied by a partially-separable model of dynamic potential outcomes. Certain MATEs are identified without direct restrictions on potential outcome persistence. The key effect homogeneity assumption permits extrapolation across observably-similar individuals from the many difference-in-difference comparisons embeded in mover designs and generates a large set of overidentifying restrictions. I characterize efficient MATE estimators and omnibus specification tests of these restrictions, both of which are straightforward to compute given a set of first-step propensity score estimates.

This analysis contributes to a small but growing econometric literature relaxing the canonical assumptions of  two-way fixed effect regressions, typically as applied to matched worker-firm panels. \cite{ams15}, for example, propose tests of the additivity restriction and develop a latent class model of non-random worker movement, while \cite{hlm17} leverage structural assumptions and long-run time series variation to estimate worker and firm effect ranks. Most recently, \cite{blm17} show how to accommodate discrete heterogeneity and Markovian job search patterns with certain forms of endogeneity. To my knowledge no paper has yet to study mover designs in a treatment effects framework, though in applications researchers sometimes appeal to the essential logic of parallel trends.\footnote{For example, \cite{fgw16}) state their identifying assumption as the restriction that ``trends do not vary systematically with the migrant's origin and destination.''}

This paper can also be thought to generalize classical and recent approaches to difference-in-differences estimation -- including \cite{ashcard85}, \cite{hit_97,hist_98}, \cite{abadie_05}, and \cite{atheyimbens06} -- to settings with multiple non-absorbing treatment states. A related recent literature considers the treatment effect interpretations of so-called ``event study'' designs, in which individuals select into a binary treatment over multiple time periods (\citealp{imaikim2016}; \citealp{bj_es}; \citealp{as_es}; \citealp{csa18}; De Chaisemartin and D'Haultfoeuille, 2018\nocite{fm16}). Here I focus on issues raised by selection into and out of multiple treatments over two periods, though an appendix section extends the theory to multiple-period mover designs. Further afield, the conditional homogeneity restriction I propose is similar to those used by \cite{angrist_fernandez_val:13}, \cite{angrist_rokkanen:15}, and \cite{hull15} for extrapolating treatment effects within and across quasi-experimental instrumental variable and regression discontinuity designs.

More generally, this paper builds on the long and rich panel data literature  \citep{chamberlain:1980, chamberlain:1982, chamberlain:1984, manski:1987, honore:1992, arellano2003} by allowing for certain forms of nonlinearity, non-additivity, and heterogeneity in causal response over time. Notably, the parallel trends assumptions I develop here are weaker than the ``time ignorability'' restrictions underlying recent approaches to non-separable panel identification (\citealp{hahn01}; \citealp{wooldridge05}; \citealp{chernetal13}), as the special structure of mover designs permits particular types of heterogeneous time-varying shocks.\footnote{\cite{graham_powell_2012} consider an alternative panel approach with continuous treatment variables, this theory is less relevant for mover selection among discrete unordered alternatives.} As mentioned above some of my proposed estimators also allow for persistent effects of treatment -- a feature that, as  \cite{imaikim2016} point out, is typically ruled out in panel data frameworks.

The remainder of this paper is organized as follows. The next section develops the dynamic treatment effects framework, characterizes the causal content of conventional mover regressions, and builds intuition for the main identification result. Section 3 then develops the class of two-step estimators for mover average treatment effects and discusses both specification testing and efficiency. Section 4 concludes. Future drafts of this paper will illustrate the theory using the the Medicare patient mover design of \cite{fgw16}. All proofs, along with the extension of the theory to multiple time periods, are contained in the appendix.

\section{Interpreting Mover Regressions\label{sec:theory}}

Suppose we observe a panel of individuals $i$ over time periods $t=0,\dots, T-1$, including an outcome $Y_{it}$, a vector of covariates $X_{it}$, and an individual's repeated selection $J_{it}\in\{0,\dots,J-1\}$ among $J$ possible treatments. Let $D_{ijt}=\mathbf{1}[J_{it}=j]$ be an indicator for individual $i$ choosing treatment $j$ in time $t$. A typical mover analysis estimates the regression
\begin{align}
Y_{it} &= \alpha_i + \tau_t + \sum_{j\neq 0} \beta_{j} D_{ijt} +X_{it}^\prime \gamma + \epsilon_{it}. \label{eq:moverreg}
\end{align}
Here $\alpha_i$ and $\tau_t$ denote individual and time fixed effects, while the coefficient $\beta_j$ is meant to capture the effect of treatment $j$ relative to the omitted treatment $0$. This \emph{mover regression} may be estimated just on the set of individuals with $J_{is}\neq J_{it}$ for some $s\neq t$ (who I refer to as \emph{movers}) or include other \emph{stayers} with $J_{it}=\bar{J}_i$ for all $t$. In practice researchers often exclude stayers or include mover-specific fixed effects in (\ref{eq:moverreg}) in order to weaken the identifying assumptions (\citealp{fgw16}; \citealp{chetty/hendren:15}).\footnote{Mover regressions are often written $Y_{it} = \alpha_i + \tau_t + \beta_{J_{it}} +X_{it}^\prime \gamma + \epsilon_{it}$. One treatment category is always omitted from estimation, though sometimes the estimated $\beta_j$ are recentered to capture effects relative to the average $j$.}

I use a dynamic potential outcomes framework (\citealp{robins86,robins97}) to characterize the causal interpretation of equation (1). Let $Y_{it}^{k\rightarrow j}$ denote the outcome of individual $i$ in time $t$ if she were to select treatment $j$ in that period after previously following the treatment path of $k=(k_0,\dots,k_{t-1})^\prime$. These are well-defined random variables under the usual stable unit treatment value assumption (\citealp{rubin80}), which I maintain throughout. When not ambiguous I write $Y_{it}^j\equiv Y_{it}^{(J_{i0},\dots,J_{i,t-1})^\prime \rightarrow j}$ as the time-$t$ potential outcome of individual $i$ given her treatment choices $J_{is}$ for $s<t$: that is, $Y_{it}^j$ implicitly conditions on choices made in past periods. Point-in-time treatment effects relative to the omitted treatment are written $Y_{it}^j-Y_{it}^0$, and realized outcomes can be written
\begin{align}
Y_{it} =& Y_{it}^0+ \sum_{j\neq 0} (Y_{it}^j-Y_{it}^0) D_{ijt}\nonumber \\
=&Y_{it}^{\overline{0}\rightarrow 0}+ \sum_{j\neq 0} (Y_{it}^{\overline{0}\rightarrow j}-Y_{it}^{\overline{0}\rightarrow 0}) D_{ijt}\label{eq:poutcomes}\\ 
&+\sum_{k\neq 0}\left(Y_{it}^{k\rightarrow 0}-Y_{it}^{\overline{0}\rightarrow 0}
+ \sum_{j\neq 0} (Y_{it}^{k\rightarrow j}-Y_{it}^{\overline{0}\rightarrow j}-(Y_{it}^{k\rightarrow 0}-Y_{it}^{\overline{0}\rightarrow 0})) D_{ijt}\right)\prod_{s<t}D_{ik_{s}s},\nonumber 
\end{align}
where $\overline{0}$ denotes a conforming vector of zeros. The first two terms of equation (\ref{eq:poutcomes}) reflect potential outcomes at time $t$ if an individual had stayed in treatment $0$ in all previous periods $s<t$, while the third term captures differences in outcomes at time $t$ arising from different treatment histories.

A comparison of equations (\ref{eq:moverreg}) and (\ref{eq:poutcomes}) suggests a set of sufficient conditions for mover regression estimates to be causally interpretable:
\begin{description}
\item [{Assumption}] IO (\emph{Impersistent outcomes}): For all $j$, $t$, and $k$, $P(Y_{it}^{k\rightarrow j}=Y_{it}^{\overline{0}\rightarrow j})=1$
\item [{Assumption}] CE (\emph{Constant effects}): For all $j$ and $t$, there exists $\overline{\beta}^j$ such that $P(Y_{it}^{j}-Y_{it}^{0}=\overline{\beta}^j)=1$
\item [{Assumption}] CO (\emph{Conditional orthogonality}): $E[D_{ijt}\epsilon_{it}]=0$ for each $j$, where $\epsilon_{it}$ denotes the residual from the population projection of $Y_{it}^{0}$ on $X_{it}$ and individual and time effects
\end{description}
Potential outcomes are impersistent under Assumption IO in that they only depend on the contemporaneous treatment status and not on previous treatment choices. When this is the case the last term of equation (\ref{eq:poutcomes}) is ignorable. When furthermore each of the period-specific treatment effects are constant across individuals (Assumption CE), we may write $Y_{it}=Y_{it}^0+\sum_{j\neq 0}\overline{\beta}^j D_{ijt}$. The regression coefficients $\beta_j$ then coincide with the causal effects $\overline{\beta}^j$ if we can decompose $Y_{it}^0 = \alpha_i + \tau_t + X_{it}^\prime \gamma + \epsilon _{it}$, with $\epsilon_{it}$ orthogonal to the vector of treatment choices (Assumption CO).\footnote{Sometimes mover regressions are motivated by a stronger conditional independence assumption, along with implicit impersistence and constant effects assumptions: see, e.g., equations (2) and (4) in \cite{ams15}.}  

Assumptions IO, CE, and CO are straightforward to state mathematically and may, as in \cite{fgw16}, stem from an underlying economic model. Nevertheless, they may also prove strong and difficult to evaluate in practice: researchers might be reluctant to rule out any forms of outcome persistence or treatment effect heterogeneity, or find it challenging to assess the appropriateness of the conditional orthogonality restriction in different settings. Instead, researchers often motivate mover regressions with claims on the comparability of outcome trends across different types of movers, and validate their estimates with the kinds of pre- and post-trend analyses typically associated with difference-in-difference designs.\footnote{See, e.g., Figure 5 in \cite{chk_2013}, Table 4 in \cite{jackson13}, and Figure 6 in \cite{fgw16}.} I therefore next consider what mover regressions identify when Assumptions IO and CE are relaxed and when Assumption CO is replaced with a quasi-experimental parallel trends assumption. To start simply and build intuition for later identification results, I first consider mover regressions with only two treatment states. 

\subsection{Binary Treatment Mover Regressions}\label{sec:twobytwo}

Suppose there are only time periods ($T=2$). Then the mover treatment coefficients are equivalently defined by a first-differenced regression,
\begin{align}
\Delta Y_{i} &= \tau + \sum_{j\neq 0} \beta_{j} \Delta D_{ij} + \Delta X_{i}^\prime \gamma + \Delta\epsilon_{i},
\end{align}
where $\Delta V_{i} = V_{i1}-V_{i0}$ denotes the first-difference operator applied to variable $V_{it}$ and $\tau=\tau_1-\tau_0$. When furthermore treatment is binary $(J=2)$ and there are no added covariates ($X_{it}=0$), we have a simple algebraic expression for the single mover regression coefficient $\beta_1$:
\begin{description}
\item [{Lemma}] 1: If $T=J=2$ and $X_{it}=0$, the mover regression coefficient equals
\begin{align}
\beta_1 =& (E[\Delta Y_i\mid\Delta D_{i1}=1]-E[\Delta Y_i\mid\Delta D_{i1}=0])\omega\label{eq:Lemma1} \\
&+(E[\Delta Y_i\mid\Delta D_{i1}=0]-E[\Delta Y_i\mid\Delta D_{i1}=-1])(1-\omega),\nonumber
\end{align}
where $\omega\in [0,1]$ is a function of $P(\Delta D_{i1}=1)$ and $P(\Delta D_{i1}=-1)$. When $P(\Delta D_{i1}=0)=0$, moreover, $\omega=1/2$ and
\begin{align}
\beta_1 =& (E[\Delta Y_i\mid\Delta D_{i1}=1]-E[\Delta Y_i\mid\Delta D_{i1}=-1])\omega.\label{eq:Lemma1b}
\end{align}
\end{description}

The proof of Lemma 1 uses omitted-variables bias algebra to write $\beta_1$ as a linear combination of the coefficients from a saturated model for $E[\Delta Y_i\mid \Delta D_{i1}]$, identifying mean outcome growth among stayers (with $\Delta D_{i1}=0$), those who move out of treatment 1 (with $\Delta D_{i1}=-1$), and those who move into treatment 1 (with $\Delta D_{i1}=1$). Lemma 1 thus shows that the simplest mover regression identifies a convex average of outcome growth comparisons between movers and stayers, across the two mover types. In the special case of no stayers this expression simplifies to the single comparison (\ref{eq:Lemma1b}) across the two mover groups.

Using Lemma 1, it is straightforward to show that a restriction on average outcome persistence, combined with a standard parallel trends assumption, renders binary treatment mover regressions causally interpretable:
\begin{description}
\item [{Proposition}] 1: If $T=J=2$, $X_{it}=0$, and for each $j\in\{0,1\}$ potential outcomes satisfy
\begin{align}
E[Y_{i1}^{(1-j)\rightarrow j}\mid \Delta D_{ij}=1]&=E[Y_{i1}^{j\rightarrow j}\mid \Delta D_{ij}=1]\label{eq:prop1_impa}
\end{align}
and
\begin{align}
E[Y_{i1}^{j\rightarrow j}-Y_{i0}^j\mid D_{ij0}D_{ij1}=1]&=E[Y_{i1}^{j\rightarrow j}-Y_{i0}^j\mid \Delta D_{ij}=1]\label{eq:prop1_ptsa}
 \\
&=E[Y_{i1}^{j\rightarrow j}-Y_{i0}^j\mid \Delta D_{ij}=-1],\label{eq:prop1_ptsb}
\end{align}
then the mover regression coefficient identifies
\begin{align}
\beta_{1}=&\sum_{t\in\{0,1\}} \sum_{d\in\{-1,1\}}E[Y_{it}^1-Y_{it}^0\mid\Delta D_{i1}=d]\hspace{0.1cm}\omega_{td},\label{eq:prop1}
\end{align}
where $\omega_{td}\ge 0$ is a function of the distribution of $(D_{i10},D_{i11})^\prime$ and $\sum\limits_{t\in\{0,1\}} \sum\limits_{d\in\{-1,1\}}\omega_{td}=1$. 
\end{description}

In words, Proposition 1 states that the binary treatment mover regression identifies a convex combination of average treatment effects, across time and the two mover groups, under two assumptions. First, equation (\ref{eq:prop1_impa}) requires individuals that move into each treatment to have, on average, the same time-$1$ outcome as if they had always been there (an impersistence assumption, weakening Assumption IO). Second, equations (\ref{eq:prop1_ptsa})-(\ref{eq:prop1_ptsb}) state that -- conditional on an individual being in treatment $j$ at any point -- the potential outcomes for different types of movers and stayers would have followed the same average growth path in the absence of a move (a parallel trends assumption, weakening Assumption CO). Note that Proposition 1 does not directly restrict treatment effect heterogeneity, relaxing Assumption CE. 

Intuition for the link between equation (\ref{eq:Lemma1}) and equation (\ref{eq:prop1}) comes from classic difference-in-differences logic. Under parallel trends, the difference in outcome growth rates between those moving into treatment $1$ at $t=1$ (with $\Delta D_{i1}=1$) and treatment $0$ stayers (with $D_{i00}=D_{i01}=1$) identifies the average time-$1$ treatment effect of the former group,
\begin{align}
&E[\Delta Y_i\mid \Delta D_{i1}=1] - E[\Delta Y_i\mid \Delta D_{i1}=0, D_{i00}=1]= E[Y_{i1}^1-Y_{i1}^0\mid \Delta D_{i1}=1],
\end{align}
while the average time-$1$ treatment effect for the other mover group (with $\Delta D_{i1}=-1$) is identified by subtracting their outcome
growth from that of the other stayer group (with $D_{i10}=D_{i11}=1$):
\begin{align}
E[\Delta Y_i\mid\Delta D_{i1}=0, D_{i10}=1 ] - E[\Delta Y_i\mid \Delta D_{i1}=-1]  = E[Y_{i1}^1-Y_{i1}^0\mid \Delta D_{i1}=-1].
\end{align}

\noindent Similarly, when the assumptions of Proposition 1 hold, comparisons of outcome growth among (i) $\Delta D_{i1}=1$ movers and treatment $1$ stayers and (ii) treatment $0$ stayers and $\Delta D_{i1}=-1$ movers identify average time-$0$ treatment effects. Thus each of the two terms in Lemma 1 can be written as a weighted average of difference-in-difference comparisons identifying average causal effects under the assumptions.

Figure 1(a) summarizes the four difference-in-difference comparisons combined in the simple mover regression. Outcome growth contrasts within the dark- and light-colored groups identify average time-$1$ effects under the assumptions of Proposition 1, while comparisons within the dashed- and solid-lines group identify average time-$0$ effects. Interestingly, only the parallel trends assumption is needed to identify the former, whereas the impersistence restriction is also used to ensure the time-$0$ comparisons are causally interpretable. This is because time-$0$ outcomes can always be ``differenced off'' when comparing within the light- dark-colored groups of Figure 1(a), as in standard difference-in-differences, but the time-$1$ outcomes in the ``reverse'' difference-in-differences of the dashed- and solid groups are not comparable when potential outcomes systematically persist. 

The appendix proof to Proposition 1 shows that the weights aggregating these difference-in-difference comparisons depend on the population shares of different mover and stayer types. Clearly if there are no movers with $\Delta D_{i1}=-1$ then $\beta_1$ represents a weighted average of causal effects for movers with $\Delta D_{i1}=1$, and vice-versa. How much weight is placed on effects from time $0$ versus time $1$ moreover depends on the proportion of stayers in each treatment: at the extremes if there are no stayers at either $d=0$ or $d=1$, then the mover regression weights together time-$d$ effects for movers with $\Delta D_{i1}=1$ and time-$(1-d)$ effects for movers with $\Delta D_{i1}=-1$. Thus when there are no initially-treated individuals, the mover regression identifies $E[Y_{i1}^1-Y_{i1}^0\mid \Delta D_{i1}=1]=E[Y_{i1}^1-Y_{i1}^0\mid D_{i11}=1]$, the average treatment effect on the treated, as in a standard difference-in-difference design.

Finally, it is worth exploring the special situation in which there are no stayers of either type. This follows from the second part of Lemma 1:
\begin{description}
\item [{Corollary to Proposition}] 1: Suppose $P(\Delta D_{i1}=0)=0$ and the assumptions of Proposition 1 hold. Then the mover regression coefficient identifies
\begin{align}
\beta_{1}&=\frac{1}{2}(E[Y_{i1}^1-Y_{i1}^0\mid \Delta D_{i1}=1]+E[Y_{i0}^1-Y_{i0}^0\mid \Delta D_{i1}=-1])\label{eq:cprop1a}\\
&=\frac{1}{2}(E[Y_{i0}^1-Y_{i0}^0\mid \Delta D_{i1}=1]+E[Y_{i1}^1-Y_{i1}^0\mid \Delta D_{i1}=-1]).\label{eq:cprop1b}
\end{align}
\end{description}
As mentioned, researchers may exclude stayers from a mover regression in order to weaken the identifying assumptions: here note that whenever the parallel trends assumption in Proposition is satisfied with stayers it is also satisfied when they are excluded. The corollary shows that without stayers the simple mover regression identifies a simply-weighted average of mover treatment effects across time. Interestingly, this estimand can be expressed in two ways: as the average of time-$t$ effects for movers into treatment 1 and time-$(1-t)$ effects for movers out of treatment 1, for either $t=0$ or $t=1$.
The equivalence of (\ref{eq:cprop1a}) and (\ref{eq:cprop1b}) is an algebraic consequence of imposing parallel trends for both treatments.

A pessimistic interpretation of Proposition 1 and its corollary is that mover regression coefficients may be difficult to interpret even in the binary treatment case. That is, two mover experiments with the same joint distribution of causal effects and treatment choices (and thus the same average treatment effects for different types of movers and stayers) may produce different regression coefficients, depending on the marginal distribution of initial treatment. Researchers interested in the external validity of mover regression estimates, or in comparing estimates across different experiments, may view this as an important limitation.\footnote{\cite{yitzhaki96} argues this point for regression-weighted averages of causal parameters in general; see also \cite{angrist98} and, more recently, \cite{BoFE18}.} Nevertheless, the above discussion suggests this limitation can be easily overcome: when the assumptions of Proposition 1 hold, a researcher wishing to estimate a time-specific average causal effect could construct the relevant difference-in-difference comparisons in Figure 1(a) and weight them together as they please. This reweighted difference-in-difference logic is at the core of the general strategy for identifying mover average treatment effects in Section 3. Before formalizing the strategy, however, we first consider additional complications arising from mover designs with multiple unordered treatments. 

\subsection{Multiple-treatment Mover Regressions}\label{sec:twobytwo}

In practice, mover regressions involve choices across many different treatments. \cite{fgw16}, for example, estimate models with $305$ healthcare market treatments, while \cite{chk_2013} study worker movement across over a million German firms. It may be reasonable to expect that the basic difference-in-difference logic extends to multiple-treatment regressions, so that they again capture some weighted average of causal effects under weak quasi-experimental restrictions. The following result, however, shows that this is not the case:

\begin{description}
\item [{Proposition}] 2: Suppose $T=2$, $X_{it}=0$, and $J>2$. Then even if both Assumption IO and the parallel trends assumption in Proposition 1 hold for all treatments $j$, the mover regression coefficients need not identify weighted averages of individual treatment effects.
\end{description}
\vspace{0.02cm}

The appendix proof of Proposition 2 shows that even with strongly impersistent outcomes and conventionally parallel trends, the multiple treatment coefficients in equation (\ref{eq:moverreg}) combine a set of average treatment effects with a set of non-causal terms. The latter are of the form 
\begin{align}
E[Y^{j}_{i1}-Y^j_{i0}\mid D_{ij0}D_{ij1}=1]-E[Y^{k}_{i1}-Y^k_{i0}\mid D_{ik0}D_{ik1}=1]\label{eq:badmulti1}
\end{align}
and 
\begin{align}
E[Y_{it}^j-Y_{it}^0\mid D_{i00}D_{ij1}=1]+E[Y_{is}^k-Y_{is}^j\mid D_{ij0}D_{ik1}=1]-E[Y_{iu}^k-Y_{iu}^0\mid D_{i00}D_{ik1}=1]\label{eq:badmulti2},
\end{align}
for treatments $j>0$ and $k\neq j$ and for time periods $t$, $s$, and $u$. These capture, respectively, differences in trends among stayers in different treatment states and combinations of causal effects for movers across different treatments and times. Although each of these comparisons are ignorable in the canonical additively-separable and constant-effects model (or imposed directly by Assumption CO and CE), in general they are non-zero under the parallel trends assumption, rendering the mover coefficients causally uninterpretable. 

At first this shortcoming of multi-treatment mover regressions may appear a puzzle. After all additional treatment choices simply yield more difference-in-difference comparisons, each of which are causally interpretable under parallel trends and Assumption IO. The issue, as with the weighting scheme of Proposition 1, is that the mover regression combines the set of quasi-experiments in a way that is sensible when the canonical mover model is correctly specified, but need not be when there are heterogeneous treatment effects or time-varying shocks.

As a simple example of the issue, consider a three-treatment design with the two mover and stayer groups illustrated in Figure 1(b). Following the logic of the previous subsection, there are two time-$1$ average causal effects identified under parallel trends: the effect of treatment $1$ relative to treatment $0$ for movers from $0$ to $1$ and the effect of treatment $2$ relative to treatment $1$ for movers from $1$ to $2$. The average time-$0$ effect of treatment $1$ relative to treatment $0$ is moreover identified for movers from $0$ to $1$ if we add an outcome impersistence condition.  As in Figure 1(a), the difference-in-difference comparisons identifying each of these effects are given by contrasts within similarly-colored and similarly-patterned line groups.

How does the mover regression combine these effects? Using formulas from the proof to Proposition 2, we can show that the treatment coefficients satisfy
\begin{align}
\beta_1 =& E[Y_{i1}^1-Y_{i1}^0\mid \Delta D_{i1}=1]p_0 + E[Y_{i0}^1-Y_{i0}^0\mid \Delta D_{i1}=1](1-p_0)\\
\beta_2 =&  E[Y_{i0}^1-Y_{i0}^0\mid \Delta D_{i1}=1]+E[Y_{i1}^2-Y_{i1}^1\mid \Delta D_{i2}=1]\label{eq:bad_beta2}\\
&+2p_0\left(E[Y_{i1}^1-Y_{i0}^1\mid D_{i10}D_{i11}=1]-E[Y_{i1}^0-Y_{i0}^0\mid D_{i00}D_{i01}=1]\right),\nonumber 
\end{align}
where $p_0$ denotes the proportion of stayers in treatment $0$. Thus, while the treatment-$1$ mover regression coefficient 
continues to identify a convex average of time-$0$ and time-$1$ treatment effects, the coefficient on treatment $2$ is not 
causal. Lacking any difference-in-difference comparison identifying the effect of treatment $2$ relative to treatment $0$, the mover regression sums the first two causal effects in equation (\ref{eq:bad_beta2}). Under a constant effects assumption this is sensible, since then
\begin{align}
E[Y_{i0}^1-Y_{i0}^0\mid \Delta D_{i1}=1]+E[Y_{i1}^2-Y_{i1}^1\mid \Delta D_{i2}=1]&=\overline{\beta}^1 + (\overline{\beta}^2-\overline{\beta}^1) = \overline{\beta}^2\nonumber \\
&=E[Y_{i1}^2-Y_{i1}^0\mid \Delta D_{i2}=1],\label{eq:cfx_extrap}
\end{align}
though with meaningful treatment effect heterogeneity this sum need not be causally interpretable. Equation (\ref{eq:bad_beta2}) moreover includes a term capturing the difference in outcome growth rates for the two stayer groups. This is again sensible under Assumptions CE and CO, since then
\begin{align}
E[Y_{i1}^1-Y_{i0}^1\mid D_{i10}D_{i11}=1]-E[Y_{i1}^0-Y_{i0}^0\mid D_{i00}D_{i01}=1]=&E[\overline{\beta}^1+Y_{i1}^0-\overline{\beta}^1 +Y_{i0}^0\mid D_{i10}D_{i11}=1]\nonumber\\
&- E[Y_{i1}^0-Y_{i0}^0\mid D_{i00}D_{i01}=1]\nonumber\\
&=0,\label{eq:trend_extrap}
\end{align}
though in general this term need not be zero. Note that the left-hand side of equation (\ref{eq:cfx_extrap}) is an example of equation (\ref{eq:badmulti2}), where the third undefined term is arbitrarily set to zero, while the left-hand side of equation (\ref{eq:trend_extrap}) is an example of equation (\ref{eq:badmulti1}).

A researcher pessimistic about the weighting scheme in Proposition 1 may therefore have even more cause for pessimism with multiple-treatment mover regressions. Despite the availability of multiple difference-in-difference comparisons, conventional mover analyses need not even have a weighted causal effect interpretation when $J>2$. Nevertheless, as with the binary treatment case, one could imagine individually extracting and more sensibly combining the component difference-in-difference quasi-experiments to overcome the limitations of mover regressions. This combination may involve extrapolating some causal effects from others, just as the above regression example does in order to identify the average effect of treatment 2 relative to treatment 0 under constant effects. Weaker extrapolations, however, may only combine difference-in-difference experiments capturing treatment effects from the same time period and for observably-similar movers, weakening the constant effects assumption. I next develop a class of two-step estimators that enact this logic.

\section{Estimating Mover Average Treatment Effects\label{sec:MATEs}}

In general there may be many combinations of heterogeneous treatment effects that are of interest in a mover design. To discipline the initial theoretical approach I focus on mover average treatment effects, defined for each treatment $j>0$ as 
\begin{align}
MATE_{jt}=E[Y_{it}^j-Y_{it}^0\mid \Delta D_i \neq 0],
\end{align}
where $\Delta D_i$ is a vector collecting the set of $\Delta D_{ij}$. Here $MATE_{jt}$ captures the average time-$t$ effect of treatment $j$ relative to treatment $0$, among individuals that change treatment status.\footnote{Recall that while for simplicity we restrict attention here to two periods, the appendix generalizes what follows to the multiple period case. For this I define $\Delta D_i$ as a vector collecting the set of $D_{ijt}-D_{ijs}$ for all $t\neq s$, so that $MATE_{jt}$ captures the average effect for individuals moving at any point.} That mover designs tend to reveal effects on mobile individuals is often implicit in applications, just as is that standard difference-in-difference estimation tends to capture average effects for those who become treated (Abadie, 2005). Of course, if there are no stayers in the study population the MATEs become average treatment effects (ATEs). 

\subsection{Identifying Assumptions}

The key quasi-experimental assumption I leverage is that of conditional parallel trends, which generalizes the trend restrictions in Proposition 1. Formally, with  $X_i$ denoting a vector of controls (including, perhaps, some elements of the $X_{i0}$ and $X_{i1}$ from the mover regression and other time-invariant observables), consider
\begin{description}
\item [{Assumption}] CPT (\emph{Conditional parallel trends}): For each treatment $j$ and $x$ in the support of $X_i$, 
\begin{align}
E[Y_{i1}^{j\rightarrow j}-Y_{i0}^j\mid D_{ij0}D_{ij1}=1,X_i=x]&=E[Y_{i1}^{j\rightarrow j}-Y_{i0}^j\mid \Delta D_{ij}=1,X_i=x]\\
&=E[Y_{i1}^{j\rightarrow j}-Y_{i0}^j\mid \Delta D_{ij}=-1,X_i=x].
\end{align}
\end{description}
Under Assumption CPT, the average treatment-$j$ outcomes for different types of movers into or out of $j$ and stayers at $j$ would have followed parallel paths if not for the move, conditional on the controls in $X_i$. In many settings it may be plausible that an individual's treatment selection is only driven by potential outcome dynamics through a set of contemporaneous or lagged observables, as with the famous ``Ashenfelter dip'' of  pre-treatment income for those entering job training programs \citep{ashenfelterdip, ashcard85}. Clearly, the parallel trends assumption in Proposition 1 is a special case of Assumption CPT, for which $X_{i}=0$. It is also straightforward to verify that the ``time ignorability'' identifying assumptions developed in the recent literature on non-separable panel models (e.g.  \cite{chernetal13}) imply Assumption CPT, but are not implied by it.\footnote{\cite{chernetal13} consider models of the form $Y_{it}^{k\rightarrow j}=g(j,X_i,\alpha_i,\epsilon_{it})$, where the distribution of $\epsilon_{it}$ does not depend on $t$ given $(\alpha_i,J_{i0},J_{i1},X_i)$. Then $E[Y_{i1}^{j\rightarrow j}-Y_{i0}^j\mid J_{i0},J_{i1},X_i]=0$, satisfying Assumption CPT. Identification here allows for heterogeneous time-varying shocks by leveraging the particular structure of mover designs.}

As with multiple-treatment mover regressions, combining causal effects across many difference-in-difference experiments requires further homogeneity restrictions. Here I adopt a weaker assumption than the canonical model: that mover treatment effects are on average comparable, conditional on observables:
\begin{description}
\item [{Assumption}] CEH (\emph{Conditional effect homogeneity}): For each period $t$, treatments $j$ and $k$,  and $x$ in the support of $X_i$, $E[Y_{it}^j-Y_{it}^k\mid \Delta D_{i}= d,X_i=x]$ does not depend on $d\neq 0$.
\end{description}
Restrictions on the conditional heterogeneity of average causal effects have been previously used in the treatment effects literature to extrapolate within and across different quasi-experiments  \citep{angrist_fernandez_val:13,angrist_rokkanen:15,hull15}. Here Assumption CEH states that differences in average causal effects for movers with different origins-destination pairs are driven only by the set of observed contemporaneous or lagged controls in $X_i$. In applications researchers sometimes gauge mover effect homogeneity by tests of outcome trend symmetry (e.g. \cite{chk_2013}); Assumption CEH can thus be thought to relax the assumptions motivating such tests. 

Finally, as in the previous section, I use a restriction on average potential outcome persistence to identify time-$0$ causal effects from ``reverse'' difference-in-difference quasi experiments. Again this can be made conditional on observables:
\begin{description}
\item [{Assumption}] COI (\emph{Conditional outcome impersistence}): For all treatments $(j,k)$ and $x\in Supp(X_i)$, 
\begin{align}
E[Y_{i1}^{k\rightarrow j}\mid D_{ik0}D_{ij1}=1,X_i=x]=&E[Y_{i1}^{j\rightarrow j}\mid D_{ik0}D_{ij1}=1,X_i=x].
\end{align}
\end{description}
Under Assumption COI, the mean outcome for movers into each treatment $j$ from each treatment $k$ is the same as it would be if the movers
had always chosen $j$, conditional on the controls in $X_i$. Thus any potential for persistence in average outcomes for movers must be driven by time-varying or invariant observables, relaxing the usual panel data restriction of complete outcome impersistence (\citealp{imaikim2016}). 

Together, these three assumptions relax those of standard mover regressions. In particular Assumptions CPT and CEH are implied by a partially-separable model of dynamic potential outcomes,
\begin{align}
Y_{i0}^{j} &= \alpha_i + \beta_{j0}(X_{i})+\epsilon_{i0}\\
Y_{i1}^{k\rightarrow j} &= \alpha_i + \beta_{j1}(X_{i})+\epsilon_{ik1},
\end{align}
where $\epsilon_{ij1}-\epsilon_{i0}$ is mean-independent of treatment choices conditional on the controls and on $i$ being a $j$-mover or $j$-stayer (that is, of $(J_{i0},J_{i1})$ conditional on $X_i$ and $D_{ij0}+D_{ij1}\neq 0$). As shown below, with enough stayers this model permits identification of time-$1$ MATEs, here taking the form 
\begin{align}
MATE_{jt}=E[\beta_{jt}(X_i)-\beta_{0t}(X_i)\mid \Delta D_i \neq 0].
\end{align}
Estimating time-$0$ MATEs or omitting stayers will additionally require Assumption COI, which here restricts $E[\epsilon_{ik1}\mid J_{i0},J_{i1},X_i]=E[\epsilon_{im1}\mid J_{i0},J_{i1},X_i]$ for all $k,m$. In contrast, the mover regression assumptions IO, CE, and CO are satisfied when the $\epsilon_{ikt}$ do not depend on $k$ and are mean-independent of $(J_{i0},J_{i1})$, and when $\beta_{jt}(X_i)$ is additively-separable in treatment, time, and a fixed linear combination of the controls. The following identification results thus allow for both time-varying shocks and treatment effect heterogeneity to vary flexibly with observables.

\subsection{Identification with and without Stayers}

With the three key assumptions in hand, I next establish MATE identification in two salient cases.  First, I suppose a researcher is willing to assume potential outcome trends are conditionally comparable between movers and stayers, and that stayers are sufficiently dispersed to make feasible a set of conditional difference-in-difference comparisons linking treatments $j$ and $0$. Enumerating this set  requires some additional notation; I let $C_j$ denote the set of all variable-length n-tuples $C_{j \ell}=(c_{j \ell0},\dots,c_{j \ell M_{j \ell}})$, where $M_{j\ell}+1$ denotes the length of $C_{j\ell}$ and where $c_{j\ell m} \in \{0,\dots,J-1\}$, $c_{j\ell 0}=0$, $c_{j\ell M_{j\ell}}= j$, and $c_{j\ell m}\neq c_{j\ell n}$ for all $m\neq n$. Thus $C_j$ collects all of the $\ell$ possible paths (or \emph{chains}) from treatment $0$ to treatment $j$ via other treatment states, where no intermediate treatment is included as a link in the chain more than once. We then have the following result:
\begin{description}
\item [{Proposition}] 3: Suppose $T=2$, Assumptions CPT and CEH hold, and that for treatment $j$ and period $t$ there exists a chain $C_{j\ell}$ such that, for each $m=1,\dots,M_{j\ell}$ and all $x$ in the support of $X_i$ either (i) $P(D_{i,m-1,1-t}D_{imt}=1\mid X_i=x)>0$ and $P(D_{i,m-1,0}D_{i,m-1,1}=1\mid X_i=x)>0$ or (ii) $P(D_{im,1-t}D_{i,m-1,t}=1\mid X_i=x)>0$ and $P(D_{im0}D_{im1}=1\mid X_i=x)>0$. Then, for $t=1$,
\begin{align}
MATE_{jt} &= E\left[\Delta Y_i \left(\sum_{m=1}^{M_{j\ell}}(w_{j\ell m}\rho_{it}^{c_{j\ell,m-1},c_{j\ell m}}+(1-w_{j\ell m})(-\rho_{it}^{c_{j\ell m},c_{j\ell,m-1}}))\right)\right],\label{eq:mate0}
\end{align}
where the $w_{j\ell m}$ are constants such that $w_{j \ell m} = 0$ if (i) fails for $m$ and $w_{j \ell m} = 1$ if (ii) fails for $m$, and where
\begin{align}
\rho_{it}^{c,d}=(-1)^t D_{ic,1-t}\frac{D_{ict}E[D_{idt}D_{ic,1-t}\mid X_{i}]-D_{idt}E[D_{ict}D_{ic,1-t}\mid X_{i}]}{E[D_{ict}D_{ic,1-t}\mid X_{i}]E[D_{idt}D_{ic,1-t}\mid X_{i}]}\frac{P(\Delta D_i\neq 0 \mid X_i)}{P(\Delta D_i\neq 0)}.\label{eq:mate0_end}
\end{align} 
If moreover Assumption COI holds, equations (\ref{eq:mate0}) and (\ref{eq:mate0_end}) also hold for $t=0$.
\end{description}

In words, Proposition 3 states that the time-$1$ mover average treatment effect for treatment $j$ is identified by a particular weighted average of outcome growth $\Delta Y_i$ under conditional parallel trends and effect homogeneity, and when there exists a set of difference-in-difference comparisons linking treatment $j$ to the reference treatment $0$. This set is given by the chain $C_{j\ell}=(0,c_{j\ell1},\dots,c_{j\ell,M_{j\ell}-1},j)$, where between any two links $c_{j\ell,m-1}$ and $c_{j\ell m}$ there exist either (i) movers from treatment $c_{j\ell,m-1}$ into treatment $c_{j\ell m}$ and stayers at treatment $c_{j\ell,m-1}$ or (ii) movers from treatment $c_{j\ell m}$ into treatment $c_{j\ell,m-1}$ and stayers at treatment $c_{j\ell m}$, conditional on the controls. The proof to Proposition 3 shows that under these assumptions the $m$-specific terms in the weighting scheme (\ref{eq:mate0}) identify $E[Y_{i1}^{c_{j\ell,m}}-Y_{i1}^{c_{j\ell,m-1}}\mid \Delta D_i\neq 0]$, so that summing over the links of the chain identifies $E[Y_{i1}^{j}-Y_{i1}^{0}\mid \Delta D_i\neq 0]=MATE_{j1}$. An analogous result follows for $t=0$ effects when potential outcomes are conditionally impersistent. 

Unpacking this result further, note that each summand of the weighting scheme (\ref{eq:mate0}) is in turn a linear combination of two terms given by equation (\ref{eq:mate0_end}). The $\rho_{it}^{c,d}$ depend on the conditional frequency of different groups of movers and stayers given the controls, as summarized by the  propensity scores $E[D_{idt}D_{ics}\mid X_i]$ and $P(\Delta D_i\mid X_i)$. The appendix proof shows that weighting $\Delta Y_i$ by $\rho_{it}^{c,d}$ and $-\rho_{it}^{d,c}$ replicates and averages together conditional difference-in-difference comparisons between the different mover and stayer groups illustrated in Figure 1(a). For example,
\begin{align}
E[\Delta Y_i \rho_{i1}^{c,d}\mid X_i] &= (E[\Delta Y_i \mid D_{ic0}D_{id1}=1,X_i]-E[\Delta Y_i \mid D_{ic0}D_{ic1}=1,X_i])\frac{P(\Delta D_i\neq 0\mid X_i)}{P(\Delta D_i\neq 0)},
\end{align}
which is a weighted comparison of the conditional difference in outcome growth between movers from treatments $c$ to $d$ and stayers at treatment $c$. Similarly, $E[\Delta Y_i (-\rho_{i1}^{d,c})\mid X_i]$ identifies a weighted conditional contrast of outcome growth between stayers at treatment $d$ and movers from treatments $d$ to $c$. As before when the conditional parallel trends assumption holds these identify conditional weighted average treatment effects of the respective mover groups, which are assumed to both be representative of all movers under conditional effect homogeneity. Thus any weighted average of $E[\Delta Y_i \rho_{i1}^{c,d}\mid X_i]$ and $E[\Delta Y_i (-\rho_{i1}^{d,c})\mid X_i]$ identifies $E[Y_{i1}^{c_{j\ell,m}}-Y_{i1}^{c_{j\ell,m-1}}\mid \Delta D_i\neq 0,X_i]\frac{P(\Delta D_i\neq 0\mid X_i)}{P(\Delta D_i\neq 0)}$ under the assumptions; averaging these averages over the marginal distribution of the controls then identifies $E[Y_{i1}^{c_{j\ell,m}}-Y_{i1}^{c_{j\ell,m-1}}\mid \Delta D_i\neq 0]$.

Proposition 3 can be thought to generalize Abadie (2005)'s approach to identification in difference-in-difference designs. Specifically, suppose $J=2$ and there are no movers from treatment $1$ to treatment $0$. Then there is only one chain $C=(0,1)$ satisfying (i), with $w_1=1$ and Assumption CEH satisfied trivially, and the weighting scheme identifying $MATE_{11}=E[Y_{i1}^1-Y_{i0}^0\mid D_{i1}=1]$ coincides with that of \cite{abadie_05}. 

It is also worth noting that the logic of Proposition 3 implies a weaker approach to MATE identification in binary treatment mover designs, using particular data-driven weights to avoid restricting treatment effect heterogeneity:
\begin{description}
\item [{Corollary to Proposition}] 3: Suppose $J=T=2$, Assumption CPT holds, and for period $t$ and the chain $C=(0,1)$ either condition (i) or (ii) from Proposition 3 holds. Then without imposing Assumption CEH we have, for $t=1$,
\begin{align}
MATE_{1t} &= E\left[\Delta Y_i \left(w_{it}^*\rho_{it}^{0,1}+(1-w_{it}^*)(-\rho_{it}^{1,0})\right)\right],\label{eq:mate0c}
\end{align}
where 
\begin{align}
w_{it}^*=\frac{P(D_{i1t}D_{i0,1-t}=1\mid X_{i})}{P(\Delta D_{i1}\neq 0 \mid X_i)}\label{eq:mate0c_end}
\end{align}
If moreover Assumption COI holds, equations (\ref{eq:mate0c}) and (\ref{eq:mate0c_end}) also hold for $t=0$.
\end{description}

As in Proposition 3, the weighting scheme in equation (\ref{eq:mate0c}) combines two sets of conditional difference-in-difference experiments via the $\rho_{it}^{0,1}$ and $(-\rho_{it}^{1,0})$ terms. Here however these terms are combined via weights  (\ref{eq:mate0c_end})  that are proportional to the share of movers of each type -- either out of or into treatment 1 -- in order to avoid mixing effects across the two groups. This corollary thus gives a flexible way to estimate pairwise average treatment effects in a mover design for movers across two treatments, though such effects need not be comparable across different treatments without an effect homogeneity assumption. 

To apply Proposition 3 or its corollary, a researcher must be willing to assume that any systematic differences in the potential outcome trends of movers and stayers arise from a set of observable controls. As in the previous section, we may also consider identification under a weaker parallel trends assumption in which the trends of stayers are unrestricted. The following result shows that in such cases one may still identify the average of $MATE_{jt}$ across time $t$:
\begin{description}
\item [{Proposition}] 4: Suppose $T=2$, Assumptions CPT, CEH, and COI hold, and for treatment $j$ there exists a chain $C_{j\ell}$ such that, for each $m=1,\dots,M_{j\ell}$ and all $x$ in the support of $X_i$, $P(D_{i,m-1,0}D_{im1}=1\mid X_i=x)>0$ and $P(D_{im0}D_{i,m-1,1}=1\mid X_i=x)>0$. Then
\begin{align}
\frac{1}{2}(MATE_{j0}+MATE_{j1}) &= E\left[\Delta Y_i \left(\sum_{m=1}^{M_{j\ell}}\kappa_{i}^{c_{j\ell,m-1},c_{j\ell m}}\right)\right],\label{eq:cmate0}
\end{align}
where
\begin{align}
\kappa_{i}^{c,d}=\frac{1}{2}\frac{D_{ic0}D_{id1}E[D_{id0}D_{ic1}\mid X_{i}]-D_{id0}D_{ic1}E[D_{ic0}D_{id1}\mid X_{i}]}{E[D_{id0}D_{ic1}\mid X_{i}]E[D_{ic0}D_{id1}\mid X_{i}]}\frac{P(\Delta D_i\neq 0 \mid X_i)}{P(\Delta D_i\neq 0)}.\label{eq:cmate0_end}
\end{align} 
\end{description}

Proposition 4 generalizes the basic logic of the corollary to Proposition 1, along the lines of Proposition 3. Here each summand in equation (\ref{eq:cmate0}) identifies a weighted difference-in-difference comparison between movers from treatment $c$ to treatment $d$ and movers from treatment $d$ to treatment $c$, such that, under the assumptions,
\begin{align}
E[\Delta Y_i \kappa_i^{c,d}\mid X_i]&= \frac{1}{2}(E[\Delta Y_i \mid D_{ic0}D_{id1}=1,X_i]-E[\Delta Y_i \mid D_{id0}D_{ic1}=1,X_i])\frac{P(\Delta D_i\neq 0\mid X_i)}{P(\Delta D_i\neq 0)}\nonumber\\ 
&= \frac{1}{2}(E[Y_{i0}^d-Y_{i0}^c \mid \Delta D_{i}\neq 0,X_i]+E[Y_{i1}^d-Y_{i1}^c \mid \Delta D_{i}\neq 0,X_i])\frac{P(\Delta D_i\neq 0\mid X_i)}{P(\Delta D_i\neq 0)}.
\end{align}
Similar to before, weighting and adding together these comparisons along the links of any chain $C_{j}$ from treatment $0$ to treatment $j$ thus identifies $\frac{1}{2}(MATE_{j0}+MATE_{j1})$. For this to be feasible there must exist both types of movers at each link, conditional on the controls, as no stayer variation is used. The cost of the weaker parallel trends assumption is a somewhat less informative estimand: the weighting scheme (\ref{eq:cmate0}) is not able to separately identify time-$0$ and time-$1$ effects. 

\subsection{Estimation and Testing}

As in \cite{abadie_05}, Propositions 3 and 4 suggest a straightforward two-step approach to estimating mover average treatment effects from an \emph{i.i.d.} sample of size $N$. In a first step, a researcher computes a set of propensity score estimates $\widehat{E}[D_{id0}D_{ic1}\mid X_i]$, along with the sample proportion of movers $\widehat{P}(\Delta D_i\neq 0)$. For a given chain $C_{j\ell}$, she then forms the sample analogue of either equation (\ref{eq:mate0}) or equation (\ref{eq:cmate0}), with the former also requiring specification of a vector of weights $w_{j\ell}$. When the first-step estimates are consistent, so too will be the second-step weighting estimator under the assumptions of Proposition 3 or 4. The asymptotic behavior of this estimator depends on the properties of the data-generating process and whether or not the first-step propensity score estimates are parametric. Future drafts will establish this behavior formally using standard first-order asymptotic theory (as in  Abadie (2005)); for now I take the $\sqrt{N}$-consistency and asymptotic normality of the described two-step estimators as given, in order to focus on additional conceptual issues arising from overidentification in mover designs.

When both conditions (i) and (ii) of Proposition 3 hold for some of the links of a chain $C_{j\ell}$, one faces the choice of which set of weights $w_{j\ell m}$ to use for estimating $MATE_{jt}$. Further degrees of overidentification arise when multiple $C_{j\ell}$ satisfy the assumptions of either Propositions 3 or 4: in general the number of possible chains grows rapidly with the number of institutions, though in practice the graph connecting any two institutions by difference-in-difference experiments may be sparse.\footnote{To be precise, there are $\sum_{k=0}^{J-2}k!\binom{J-2}{k}^2$ possible chains given $J$ institutions.} Testing the equality of two estimates formed by different chains or weights constitutes an omnibus specification test of the identifying assumptions which may be used to gauge the plausibility of the quasi-experimental framework.

A potentially more powerful test of this joint null hypothesis uses a conditionally-efficient estimator, optimally combining different quasi-experiments to minimize the asymptotic estimator variance. To characterize this procedure, let $\widehat{\rho}_{it}^{c,d}$ and $\widehat{\kappa}_{it}^{c,d}$ be consistent estimates of $\rho_{it}^{c,d}$ and $\kappa_{it}^{c,d}$, and let $\widehat{M}$ denote a $K\times 1$ vector collecting the set of either all positive and negative $\widehat{E}[\Delta Y_i \widehat{\rho}_{it}^{c,d}]$ or of all $\widehat{E}[\Delta Y_i \widehat{\kappa}_{it}^{c,d}]$ that are well-defined for treatments $(c,d)$, where $\widehat{E}[V_{it}]$ denotes the sample average of a variable $V_{it}$. Next, let $S_{j}$ be a $P\times K$ matrix with elements $S_{jpk}\in\{0,1\}$, such that, for all $p$, $\sum_k S_{jpk}\widehat{M}_k$ is consistent for $MATE_{jt}$ under Proposition 3 or identifies $\frac{1}{2}(MATE_{j0}+MATE_{j1})$ under Proposition 4. Then a general method of moments estimator for the target causal parameter $\beta_j$, which combines mover variation across all available chains, is given by
\begin{align}
\widehat{\beta}_j&=\arg\min_{\beta_j} (\beta_j - S_j \widehat{M})^\prime W(\beta_j - S_j \widehat{M}),\label{eq:GMM}
\end{align}
where $W$ is some $P\times P$ positive-definite weighting matrix. Given the first-step propensity scores underlying $\widehat{M}$, a conditionally-efficient estimator $\widehat{\beta}_j^*$ is obtained by setting $W=(S_j\widehat{\Omega}S_j^{\prime})^{-1}$, where $\widehat{\Omega}$ consistently estimates the asymptotic variance of $\widehat{M}$. Solving (\ref{eq:GMM}), we then have
\begin{align}
\widehat{\beta}_j^* &= \frac{\iota^\prime (S_j\widehat{\Omega}S_j^\prime)^{-1}S_j}{\iota^\prime (S_j\widehat{\Omega}S_j^\prime)^{-1}\iota}\widehat{M},\label{eq:betastar}
\end{align}
where $\iota$ denotes a $P\times 1$ vector of ones. 

This estimate of the target causal parameter (either $MATE_{jt}$ if $\widehat{M}$ collects moments involving $\widehat{\rho}_{it}^{c,d}$ or $\frac{1}{2}(MATE_{j0}+MATE_{j1})$ if $\widehat{M}$ collects moments involving $\widehat{\kappa}_{it}^{c,d}$) is an optimally-weighted combination of all relevant difference-in-difference quasi-experiments, with weights proportional to the elements of $\iota^\prime (S_j\widehat{\Omega}S_j^\prime)^{-1}S_j$. In particular, it satisfies
\begin{align}
\sqrt{N}(\hat{\beta}_j^*-\beta) \Rightarrow \mathcal{N}(0,(\iota^\prime(S_j\Omega S_j^\prime)^{-1}\iota)^{-1}),
\end{align}
where $\widehat{\Omega}\xrightarrow{p}\Omega$. As usual, an omnibus specification test based on (\ref{eq:betastar}) is given by 
\begin{align}
\widehat{T}_j &= (\beta_j^* - S_j \widehat{M})^\prime (S_j\widehat{\Omega}S_j^{\prime})^{-1}(\beta_j^* - S_j \widehat{M}),\label{eq:Test}
\end{align}
which has an asymptotic $\chi^2_{P-1}$ distribution under the joint null of the identifying assumptions. Implicitly, (\ref{eq:Test}) checks whether the same estimate of $\beta_j$ is obtained from any two difference-in-difference chains, weighting pairwise comparisons by the efficient weights.

It appears relatively straightforward to compute equations (\ref{eq:betastar}) and (\ref{eq:Test}) in practice. Estimation of the large set of mover propensity scores can be distributed over multiple computational resources, while the second-step calculation of the estimated asymptotic variance of $\widehat{M}$ and the second-step estimator (\ref{eq:betastar}) is likely to be quite simple. Future drafts of this paper will include simulations of the computational demands and finite-sample performance of these estimators, as well as an application to a real-world movers design.





\section{Conclusions\label{sec:conclusions}}

Although retaining the flavor of simpler difference-in-differences designs, quasi-experimental mover designs require additional restrictions. Mover regression estimates, while causally interpretable in the binary treatment case, fail in general to recover weighted averages of heterogeneous treatment effects under a parallel trends assumption alone. In contrast, the two-step estimators developed here accommodate
heterogeneous treatment effects and time-varying shocks, provided they are not correlated with individual movement conditional on controls. Certain mover average treatment effects are moreover identified without a direct restriction on potential outcome persistence, provided the potential trends of movers and stayers are conditionally comparable. As argued above, the computation of conditionally-efficient treatment effect estimates is likely to be light in practice, even relative to recent advances in traditional two-way fixed effect regressions \citep{ack2002, guimaraes/portugal:10, gaure13, correia2016}. 

As always, whether the assumptions considered here are more plausible than other approaches to mover designs , such as \cite{hlm17} and \cite{blm17}, will be a matter of context. The restrictions of conditional parallel trends and conditional effect homogeneity are likely to have the advantage of both being familiar to applied researchers and relatively straightforward to consider in applications. At a minimum, the quasi-experimental approach may allow researchers to verify the robustness of substantive conclusions drawn from conventional mover regressions and more recent variations.

\pagebreak{}

\section*{Figures and Tables}

\vspace{3.5cm}
\begin{figure}[!h]
\centering
\includegraphics[trim=6cm 8.32cm 4.1cm 5.9cm, clip, height=6cm]{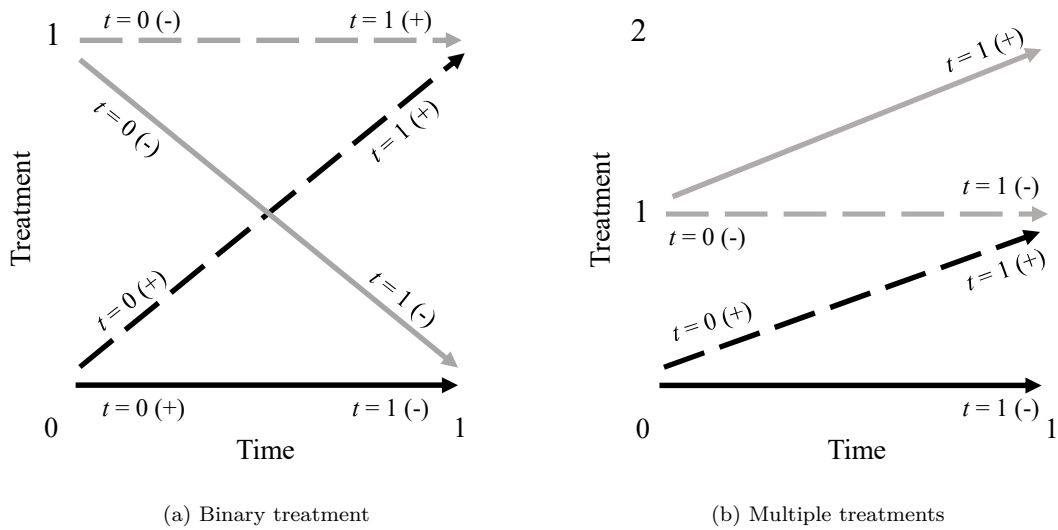} 
\subfloat[Binary treatment]{\hspace{.48\linewidth}}
\subfloat[Multiple treatments]{\hspace{.5\linewidth}}
\caption[moverfig]{Example difference-in-difference comparisons in mover designs\label{illustration}}
\vspace{0.35cm} 
\begin{minipage}{0.92\textwidth}
{\footnotesize Notes: This figure illustrates groups of movers and stayers identifying time-$t$ causal effects when outcomes are impersistent and parallel trends holds. In each panel, time-$0$ effects are identified by outcome growth contrasts within the dashed- and (in Figure 1(a)) solid-line groups, while time-$1$ effects are identified by outcome growth contrasts within the light- and dark-colored groups. The time-specific notes on each line indicate whether the outcome growth for that subgroup is to be added or subtracted. Panel (a) shows the full set of difference-in-difference comparisons with binary treatments, while panel (b) shows the comparisons for the multiple treatment example discussed in the text.}
\end{minipage}
\end{figure}

\pagebreak{}
\begin{spacing}{1.12}
\bibliographystyle{ecta}
\bibliography{twoway}

\begin{thebibliography}{49}
\newcommand{\enquote}[1]{``#1''}
\expandafter\ifx\csname natexlab\endcsname\relax\def\natexlab#1{#1}\fi

\bibitem[\protect\citeauthoryear{Abadie}{Abadie}{2005}]{abadie_05}
\textsc{Abadie, A.} (2005): \enquote{Semiparametric Difference-in-Differences
  Estimators,} \emph{Review of Economic Studies}, 72, 1--19.

\bibitem[\protect\citeauthoryear{Abowd, Creecy, and Kramarz}{Abowd
  et~al.}{2002}]{ack2002}
\textsc{Abowd, J., R.~Creecy, and F.~Kramarz} (2002): \enquote{Computing Person
  and Firm Effects Using Linked Longitudinal Employer-Employee Data,} Cornell
  University Department of Economics Unpublished Working Paper.

\bibitem[\protect\citeauthoryear{Abowd, Kramarz, and Margolis}{Abowd
  et~al.}{1999}]{akm99}
\textsc{Abowd, J.~M., F.~Kramarz, and D.~Margolis} (1999): \enquote{High Wage
  Workers and High Wage Firms,} \emph{Econometrica}, 67, 251--333.

\bibitem[\protect\citeauthoryear{Abowd, McKinney, and Schmutte}{Abowd
  et~al.}{2015}]{ams15}
\textsc{Abowd, J.~M., K.~L. McKinney, and I.~M. Schmutte} (2015):
  \enquote{Modeling Endogenous Mobility in Wage Determination,} Working Paper.

\bibitem[\protect\citeauthoryear{Abraham and Sun}{Abraham and
  Sun}{2018}]{as_es}
\textsc{Abraham, S. and L.~Sun} (2018): \enquote{Estimating Dynamic Treatment
  Effects in Event Studies,} Working Paper.

\bibitem[\protect\citeauthoryear{Allcott, Diamond, and Dub{\'e}}{Allcott
  et~al.}{2017}]{alcottetal17}
\textsc{Allcott, H., R.~Diamond, and J.-P. Dub{\'e}} (2017): \enquote{The
  Geography of Poverty and Nutrition: Food Deserts and Food Choices Across the
  United States,} Working Paper.

\bibitem[\protect\citeauthoryear{Angrist and Fernandez-Val}{Angrist and
  Fernandez-Val}{2013}]{angrist_fernandez_val:13}
\textsc{Angrist, J. and I.~Fernandez-Val} (2013): \enquote{ExtrapoLATE-ing:
  External Validity and Overidentification in the LATE Framework,}
  \emph{Advances in Economics and Econometrics: Theory and Applications, Tenth
  World Congress}, 3, 401--433.

\bibitem[\protect\citeauthoryear{Angrist and Rokkanen}{Angrist and
  Rokkanen}{2015}]{angrist_rokkanen:15}
\textsc{Angrist, J. and M.~Rokkanen} (2015): \enquote{Wanna Get Away?
  Regression DIscontinuity Estimation of Exam School Effects Away from the
  Cutoff,} \emph{Journal of the American Statistical Association}, 110,
  1331--1344.

\bibitem[\protect\citeauthoryear{Angrist}{Angrist}{1998}]{angrist98}
\textsc{Angrist, J.~D.} (1998): \enquote{Estimating the Labor Market Impact of
  Voluntary Military Service Using Social Security Data on Military
  Applicants,} \emph{Econometrica}, 66, 249--288.

\bibitem[\protect\citeauthoryear{Arellano}{Arellano}{2003}]{arellano2003}
\textsc{Arellano, M.} (2003): \emph{Panel Data Econometrics}, Oxford University
  Press.

\bibitem[\protect\citeauthoryear{Ashenfelter}{Ashenfelter}{1978}]{ashenfelterdip}
\textsc{Ashenfelter, O.} (1978): \enquote{Estimating the Effect of Training
  Programs on Earnings,} \emph{Review of Economics and Statistics}, 60, 47--57.

\bibitem[\protect\citeauthoryear{Ashenfelter and Card}{Ashenfelter and
  Card}{1985}]{ashcard85}
\textsc{Ashenfelter, O. and D.~Card} (1985): \enquote{Using the Longitudinal
  Structure of Earnings to Estimate the Effect of Training Programs,}
  \emph{Review of Economics and Statistics}, 67, 648--660.

\bibitem[\protect\citeauthoryear{Athey and Imbens}{Athey and
  Imbens}{2006}]{atheyimbens06}
\textsc{Athey, S. and G.~W. Imbens} (2006): \enquote{Identification and
  Inference in Nonlinear Difference-in-Differences Models,}
  \emph{Econometrica}, 74, 431--497.

\bibitem[\protect\citeauthoryear{Bloom, Song, Price, Guvenen, and von
  Wachter}{Bloom et~al.}{2015}]{fui2015}
\textsc{Bloom, N., J.~Song, D.~Price, F.~Guvenen, and T.~von Wachter} (2015):
  \enquote{Firming up Inequality,} \emph{NBER Working Paper. 21199}.

\bibitem[\protect\citeauthoryear{Bonhomme, Lamadon, and Manresa}{Bonhomme
  et~al.}{2017}]{blm17}
\textsc{Bonhomme, S., T.~Lamadon, and E.~Manresa} (2017): \enquote{A
  Distributional Framework for Matched Employer Employee Data,} Working Paper.

\bibitem[\protect\citeauthoryear{Borusyak and Jaravel}{Borusyak and
  Jaravel}{2016}]{bj_es}
\textsc{Borusyak, K. and X.~Jaravel} (2016): \enquote{Revisiting Event Study
  Designs, with an Application to the Estimation of the Marginal Propensity to
  Consume,} Working Paper.

\bibitem[\protect\citeauthoryear{Bronnenberg, Dub{\'e}, and
  Gentzkow}{Bronnenberg et~al.}{2012}]{bdg12}
\textsc{Bronnenberg, B.~J., J.-P. Dub{\'e}, and M.~Gentzkow} (2012):
  \enquote{The Evolution of Brand Preferences: Evidence from Consumer
  Migration,} \emph{American Economic Review}, 102, 2472--2508.

\bibitem[\protect\citeauthoryear{Callaway and Sant'Anna}{Callaway and
  Sant'Anna}{2018}]{csa18}
\textsc{Callaway, B. and P.~H.~C. Sant'Anna} (2018):
  \enquote{Difference-in-Differences with Multiple Time Periods and an
  Application on the Minimum Wage and Employment,} Working Paper.

\bibitem[\protect\citeauthoryear{Card, Heining, and Kline}{Card
  et~al.}{2013}]{chk_2013}
\textsc{Card, D., J.~Heining, and P.~Kline} (2013): \enquote{Workplace
  Heterogeneity and the Rise of West German Wage Inequality,} \emph{Quarterly
  Journal of Economics}, 128, 967--1015.

\bibitem[\protect\citeauthoryear{Chamberlain}{Chamberlain}{1980}]{chamberlain:1980}
\textsc{Chamberlain, G.} (1980): \enquote{Analysis of Covariance with
  Qualitative Data,} \emph{Review of Economic Studies}, 47, 225--238.

\bibitem[\protect\citeauthoryear{Chamberlain}{Chamberlain}{1982}]{chamberlain:1982}
---\hspace{-.1pt}---\hspace{-.1pt}--- (1982): \enquote{Multivariate Regression
  Models for Panel Data,} \emph{Journal of Econometrics}, 18, 5--46.

\bibitem[\protect\citeauthoryear{Chamberlain}{Chamberlain}{1984}]{chamberlain:1984}
---\hspace{-.1pt}---\hspace{-.1pt}--- (1984): \enquote{Panel Data,} in
  \emph{Handbook of Econometrics}, ed. by Z.~Griliches and M.~D. Intriligator,
  Elsevier, vol.~2, chap.~22, 1247--1318, 1 ed.

\bibitem[\protect\citeauthoryear{Chernozhukov, Fernandez-Val, Hahn, and
  Newey}{Chernozhukov et~al.}{2013}]{chernetal13}
\textsc{Chernozhukov, V., I.~Fernandez-Val, J.~Hahn, and W.~Newey} (2013):
  \enquote{Average and Quantile Effects in Nonseparable Panel Models,}
  \emph{Econometrica}, 81, 535--580.

\bibitem[\protect\citeauthoryear{Chetty, Friedman, and Rockoff}{Chetty
  et~al.}{2014}]{cfr:14a}
\textsc{Chetty, R., J.~Friedman, and J.~Rockoff} (2014): \enquote{Measuring the
  Impact of Teachers I: Evaluating Bias in Teacher Value-Added Estimates,}
  \emph{American Economic Review}, 104(9), 2593--2632.

\bibitem[\protect\citeauthoryear{Chetty and Hendren}{Chetty and
  Hendren}{Forthcoming}]{chetty/hendren:15}
\textsc{Chetty, R. and N.~Hendren} (Forthcoming): \enquote{The Impacts of
  Neighborhoods on Intergenerational Mobility I: Childhood Exposure Effects,}
  \emph{Quarterly Journal of Economics}.

\bibitem[\protect\citeauthoryear{Correia}{Correia}{2016}]{correia2016}
\textsc{Correia} (2016): \enquote{A Feasible Estimator for Linear Models with
  Multi-way Fixed Effects,} Working Paper.

\bibitem[\protect\citeauthoryear{de~Chaisemartin and
  D'Haultfoeuille}{de~Chaisemartin and D'Haultfoeuille}{2018}]{fm16}
\textsc{de~Chaisemartin, C. and X.~D'Haultfoeuille} (2018): \enquote{Two-way
  Fixed Effects Estimators with Heterogeneous Treatment Effects,} \emph{Working
  Paper}.

\bibitem[\protect\citeauthoryear{Finkelstein, Gentzkow, Hull, and
  Williams}{Finkelstein et~al.}{2017}]{fghw17}
\textsc{Finkelstein, A., M.~Gentzkow, P.~Hull, and H.~Williams} (2017):
  \enquote{Adjusting Risk Adjustment: Accounting for Variation in Diagnostic
  Intensity,} \emph{New England Journal of Medicine}, 376, 608--610.

\bibitem[\protect\citeauthoryear{Finkelstein, Gentzkow, and
  Williams}{Finkelstein et~al.}{2016}]{fgw16}
\textsc{Finkelstein, A., M.~Gentzkow, and H.~Williams} (2016): \enquote{Sources
  of Geographic Variation in Health Care: Evidence from Patient Migration,}
  \emph{Quarterly Journal of Economics}, 131, 1681--1726.

\bibitem[\protect\citeauthoryear{Gaure}{Gaure}{2013}]{gaure13}
\textsc{Gaure, S.} (2013): \enquote{OLS with Multiple High Dimensional Category
  Variables,} \emph{Computational Statistics and Data Analysis}, 66, 8--18.

\bibitem[\protect\citeauthoryear{Gibbons, Serrato, and Urbancic}{Gibbons
  et~al.}{2018}]{BoFE18}
\textsc{Gibbons, C.~E., J.~C.~S. Serrato, and M.~B. Urbancic} (2018):
  \enquote{Broken or Fixed Effects?} \emph{Journal of Econometric Methods},
  20170002.

\bibitem[\protect\citeauthoryear{Graham and Powell}{Graham and
  Powell}{2012}]{graham_powell_2012}
\textsc{Graham, B.~S. and J.~L. Powell} (2012): \enquote{Identification and
  Estimation of Average Partial Effects in ``Irregular'' Correlated Random
  Coefficient Panel Data Models,} \emph{Econometrica}, 80, 2105--2152.

\bibitem[\protect\citeauthoryear{Guimaraes and Portugal}{Guimaraes and
  Portugal}{2010}]{guimaraes/portugal:10}
\textsc{Guimaraes, P. and P.~Portugal} (2010): \enquote{A Simple Feasible
  Procedure to Fit Models with High-Dimensional Fixed Effects,} \emph{Stata
  Journal}, 10, 628--649.

\bibitem[\protect\citeauthoryear{Hagedorn, Law, and Manovskii}{Hagedorn
  et~al.}{2017}]{hlm17}
\textsc{Hagedorn, M., T.~H. Law, and I.~Manovskii} (2017): \enquote{Identifying
  Equilibrium Models of Labor Market Sorting,} \emph{Econometrica}, 85, 29--65.

\bibitem[\protect\citeauthoryear{Hahn}{Hahn}{2001}]{hahn01}
\textsc{Hahn, J.} (2001): \enquote{Comment: Binary Regressors in Nonlinear
  Panel-Data Models with Fixed Effects,} \emph{Journal of Business and Economic
  Statistics}, 19, 16--17.

\bibitem[\protect\citeauthoryear{Heckman, Ichimura, Smith, and Todd}{Heckman
  et~al.}{1998}]{hist_98}
\textsc{Heckman, J.~J., H.~Ichimura, J.~Smith, and P.~Todd} (1998):
  \enquote{Characterizing Selection Bias Using Expermental Data,}
  \emph{Econometrica}, 66, 1017--1098.

\bibitem[\protect\citeauthoryear{Heckman, Ichimura, and Todd}{Heckman
  et~al.}{1997}]{hit_97}
\textsc{Heckman, J.~J., H.~Ichimura, and P.~E. Todd} (1997): \enquote{Matching
  as an Evaluation Estimator: Evidence from Evaluating a Job Training
  Programme,} \emph{Review of Economic Studies}, 64, 605--654.

\bibitem[\protect\citeauthoryear{Honore}{Honore}{1992}]{honore:1992}
\textsc{Honore, B.~E.} (1992): \enquote{Trimmed Lad and Least Squares
  Estimation of Truncated and Censored Regression Models wth Fixed Effects,}
  \emph{Econometrica}, 60, 533--565.

\bibitem[\protect\citeauthoryear{Hull}{Hull}{2018}]{hull15}
\textsc{Hull, P.} (2018): \enquote{IsoLATEing: Identifying
  Counterfactual-Specific Treatment Effects with Cross-Stratum Comparisons,}
  Working Paper.

\bibitem[\protect\citeauthoryear{Imai and Kim}{Imai and
  Kim}{2016}]{imaikim2016}
\textsc{Imai, K. and I.~S. Kim} (2016): \enquote{When Should We Use Linear
  Fixed Effects Regression Models for Causal Inference with Longitudinal Data?}
  Working Paper.

\bibitem[\protect\citeauthoryear{Jackson}{Jackson}{2013}]{jackson13}
\textsc{Jackson, C.~K.} (2013): \enquote{Match Quality, Worker Productivity,
  and Worker Mobility: Direct Evidence from Teachers,} \emph{The Review of
  Economics and Statistics}, 95, 1096--1116.

\bibitem[\protect\citeauthoryear{Manski}{Manski}{1987}]{manski:1987}
\textsc{Manski, C.} (1987): \enquote{Semiparametric Analysis of Random Effects
  Linear Models from Binary Response Data,} \emph{Econometrica}, 55, 357--362.

\bibitem[\protect\citeauthoryear{Molitor}{Molitor}{2017}]{molitor17}
\textsc{Molitor, D.} (2017): \enquote{The Evolution of Physician Practice
  Styles: Evidence from Cardiologist Migration,} \emph{American Economic
  Journal: Economic Policy}, 10, 326--356.

\bibitem[\protect\citeauthoryear{Robins}{Robins}{1986}]{robins86}
\textsc{Robins, J.} (1986): \enquote{A New Approach to Causal Inference in
  Mortality Studies with a Sustained Exposure Period: Application to Control of
  the Healthy Worker Survivor Effect,} \emph{Mathematical Modelling}, 7,
  1393--1512.

\bibitem[\protect\citeauthoryear{Robins}{Robins}{1997}]{robins97}
---\hspace{-.1pt}---\hspace{-.1pt}--- (1997): \enquote{Causal Inference from
  Complex Longitudinal Data,} in \emph{Latent Variable Modeling and
  Applications to Causality: Lecture Notes in Statistics}, ed. by M.~Berkane,
  Springer Verlag, vol. 120, 69--117.

\bibitem[\protect\citeauthoryear{Rubin}{Rubin}{1980}]{rubin80}
\textsc{Rubin, D.~B.} (1980): \enquote{Randomization Analysis of Experimental
  Data: The Fisher Randomization Test Comment,} \emph{Journal of the American
  Statistical Association}, 75, 591--593.

\bibitem[\protect\citeauthoryear{Sacarny}{Sacarny}{2016}]{sacarnyjmp}
\textsc{Sacarny, A.} (2016): \enquote{Technological Diffusion Across Hospitals:
  The Case of a Revenue-Generating Practice,} Working Paper.

\bibitem[\protect\citeauthoryear{Wooldridge}{Wooldridge}{2005}]{wooldridge05}
\textsc{Wooldridge, J.~M.} (2005): \enquote{Fixed-Effects and Related
  Estimators for Correlated Random-Coefficient and Treatment-Effect Panel Data
  Models,} \emph{The Review of Economics and Statistics}, 87, 385--390.

\bibitem[\protect\citeauthoryear{Yitzhaki}{Yitzhaki}{1996}]{yitzhaki96}
\textsc{Yitzhaki, S.} (1996): \enquote{On Using Linear Regressions in Welfare
  Economics,} \emph{Journal of Business and Economic Statistics}, 14, 478--486.

\end{thebibliography}
\end{spacing}
\pagebreak{}

\section*{Appendix}

\subsubsection*{Proof of Lemma 1}
With $T=J=2$ and $X_{it}=0$, the mover regression can be written
\begin{align}
\Delta Y_i &= \tau + \beta_1 \Delta D_{i1} + \Delta \epsilon_i\nonumber \\
& = \tau + \beta_1 (\mathbf{1}[\Delta D_{i1}=1] - \mathbf{1}[\Delta D_{i1} = -1]) + \Delta \epsilon_i,
\end{align}
which is nested by the regression
\begin{align}
\Delta Y_i &=  \widetilde{\tau} + \widetilde{\beta}_1 (\mathbf{1}[\Delta D_{i1}=1] - \mathbf{1}[\Delta D_{i1} = -1]) + \widetilde{\alpha}_1\mathbf{1}[\Delta D_{i1}=-1] + \Delta \widetilde{\epsilon}_i \nonumber \\
&=  \widetilde{\tau} + \widetilde{\beta}_1 \mathbf{1}[\Delta D_{i1}=1] + (\widetilde{\alpha}_1-\widetilde{\beta}_1)\mathbf{1}[\Delta D_{it}=-1] + \Delta \widetilde{\epsilon}_i.
\end{align}
This is a saturated model for $E[\Delta Y_i \mid \Delta D_{i1}]$, with
\begin{align}
\widetilde{\beta}_1 &= E[\Delta Y_i\mid\Delta D_{i1}=1]-E[\Delta Y_i\mid\Delta D_{i1}=0]\label{eq:beta_tilde_1} \\ 
\text{and }\hspace{0.25cm}\widetilde{\alpha}_1-\widetilde{\beta}_1 & = E[\Delta Y_i\mid\Delta D_{i1}=-1]-E[\Delta Y_i\mid\Delta D_{i1}=0].\label{eq:alpha_tilde_1}
\end{align}
Thus, by usual omitted-variables bias logic,
\begin{align}
\beta_1 = \widetilde{\beta}_1 + \widetilde{\alpha}_1 \frac{Cov( \mathbf{1}[\Delta D_{i1} = -1]),\Delta D_{i1})}{Var(\Delta D_{i1})}.
\end{align}
Define 
\begin{align}
\omega &= 1+ \frac{Cov( \mathbf{1}[\Delta D_{i1} = -1]),\Delta D_{i1})}{Var(\Delta D_{i1})}\nonumber \\
&=\frac{p^+(1-p^+)+p^+p^-}{p^+(1-p^+)+p^-(1-p^-)+2p^+p^-},
\end{align}
where $p^+=P(\Delta D_{i1}=1)$ and $p^-=P(\Delta D_{i1}=-1)$. Note that $\omega \in [0,1]$ and $\omega=1/2$ when $1-p^+=p^-$. Thus, 
\begin{align}
\beta_1 = \widetilde{\beta}_1 \omega + (\widetilde{\beta}_1-\widetilde{\alpha}_1) (1-\omega).\label{eq:beta_1}
\end{align}
Combining equations (\ref{eq:beta_tilde_1}), (\ref{eq:alpha_tilde_1}), and (\ref{eq:beta_1}) completes the proof.$\hfill\square$

\newpage

\subsubsection*{Proof of Proposition 1 and Corollary}
First suppose $P(\Delta D_{i1}=0)>0$. Under the impersistence and parallel trends assumptions,
\begin{align}
&E[\Delta Y_i\mid\Delta D_{i1}=1]-E[\Delta Y_i\mid\Delta D_{i1}=0]\nonumber \\
&=\left(E[Y_{i1}-Y_{i0}\mid\Delta D_{i1}=1]-E[Y_{i1}-Y_{i0}\mid\Delta D_{i1}=0,D_{i00}=1]\right)p\nonumber\\
&\quad+\left(E[Y_{i1}-Y_{i0}\mid\Delta D_{i1}=1]-E[Y_{i1}-Y_{i0}\mid\Delta D_{i1}=0,D_{i10}=1]\right)(1-p)\nonumber \\
&=\left(E[Y_{i1}^{0\rightarrow 1}-Y_{i0}^0\mid\Delta D_{i1}=1]-E[Y_{i1}^{0\rightarrow 0}-Y_{i0}^0\mid\Delta D_{i1}=0,D_{i00}=1]\right)p\nonumber\\
&\quad+\left(E[Y_{i1}^{0\rightarrow 1}-Y_{i0}^0\mid\Delta D_{i1}=1]-E[Y_{i1}^{1\rightarrow 1}-Y_{i0}^1\mid\Delta D_{i1}=0,D_{i10}=1]\right)(1-p)\nonumber \\
&=E[Y_{i1}^1-Y_{i1}^0\mid\Delta D_{i1}=1]p+E[Y_{i0}^1-Y_{i0}^0\mid\Delta D_{i1}=1](1-p),
\end{align}
where $p=P(D_{i00}=1\mid\Delta D_{i1}=0)$. Similarly, 
\begin{align}
&E[\Delta Y_i\mid\Delta D_{i1}=0]-E[\Delta Y_i\mid\Delta D_{i1}=-1]\nonumber \\
&=\left(E[Y_{i1}-Y_{i0}\mid\Delta D_{i1}=0,D_{i00}=1]-E[Y_{i1}-Y_{i0}\mid\Delta D_{i1}=-1]\right)p\nonumber\\
&\quad+\left(E[Y_{i1}-Y_{i0}\mid\Delta D_{i1}=0,D_{i10}=1]-E[Y_{i1}-Y_{i0}\mid\Delta D_{i1}=-1]\right)(1-p)\nonumber \\
&=\left(E[Y_{i1}^{0\rightarrow 0}-Y_{i0}^0\mid\Delta D_{i1}=0,D_{i00}=1]-E[Y_{i1}^{1\rightarrow 0}-Y_{i0}^1\mid\Delta D_{i1}=-1]\right)p\nonumber\\
&\quad+\left(E[Y_{i1}^{1\rightarrow 1}-Y_{i0}^1\mid\Delta D_{i1}=0,D_{i10}=1]-E[Y_{i1}^{1\rightarrow 0}-Y_{i0}^1\mid\Delta D_{i1}=-1]\right)(1-p)\nonumber \\
&=E[Y_{i0}^1-Y_{i0}^0\mid\Delta D_{i1}=-1]p+E[Y_{i1}^1-Y_{i1}^0\mid\Delta D_{i1}=-1](1-p).
\end{align}
Substituting these expressions in to the equation for the regression coefficient in Lemma 1 gives
\begin{align}
\beta_{1}=&E[Y_{i1}^1-Y_{i1}^0\mid\Delta D_{i1}=1]p\omega+E[Y_{i1}^1-Y_{i1}^0\mid\Delta D_{i1}=-1](1-p)(1-\omega) \\
&+E[Y_{i0}^1-Y_{i0}^0\mid\Delta D_{i1}=1](1-p)\omega+E[Y_{i0}^1-Y_{i0}^0\mid\Delta D_{i1}=-1]p(1-\omega).\nonumber
\end{align}
Now consider the corollary case of $P(\Delta D_{i1}=0)=0$. From Lemma 1,
\begin{align}
\beta_{1}=&(E[Y_{i1}^1-Y_{i0}^0\mid\Delta D_{i1}=1]-E[Y_{i1}^0-Y_{i0}^1\mid\Delta D_{i1}=-1])/2\nonumber\\
=&(E[Y_{i1}^{0\rightarrow 1}-Y_{i0}^0\mid\Delta D_{i1}=1]-E[Y_{i1}^{0\rightarrow 1}-Y_{i0}^1\mid\Delta D_{i1}=1])/2\nonumber\\
&-(E[Y_{i1}^{1\rightarrow 0}-Y_{i0}^1\mid\Delta D_{i1}=-1]-E[Y_{i1}^{1\rightarrow 1}-Y_{i0}^1\mid\Delta D_{i1}=-1])/2\nonumber\\
=&(E[Y_{i0}^1-Y_{i0}^0\mid\Delta D_{i1}=1]+E[Y_{i1}^1-Y_{i1}^0\mid\Delta D_{i1}=-1])/2\nonumber \\
=&(E[Y_{i1}^{0\rightarrow 1}-Y_{i0}^0\mid\Delta D_{i1}=1]-E[Y_{i1}^{0\rightarrow 0}-Y_{i0}^0\mid\Delta D_{i1}=1])/2\nonumber\\
&-(E[Y_{i1}^{1\rightarrow 0}-Y_{i0}^1\mid\Delta D_{i1}=-1]-E[Y_{i1}^{1\rightarrow 0}-Y_{i0}^0\mid\Delta D_{i1}=-1])/2\nonumber\\
=&(E[Y_{i1}^1-Y_{i1}^0\mid\Delta D_{i1}=1]+E[Y_{i0}^1-Y_{i0}^0\mid\Delta D_{i1}=-1])/2,
\end{align}
where the second and fourth lines again follow from impersistence and parallel trends. $\hfill \square$

\newpage
\subsubsection*{Proof of Proposition 2}

The mover regression
\begin{align}
\Delta Y_i &= \tau + \sum_{j\neq 0}\beta_j  (\mathbf{1}[\Delta D_{i1}=1] - \mathbf{1}[\Delta D_{i1} = -1]) + \Delta \epsilon_i.
\end{align}
is nested by the model
\begin{align}
\Delta Y_i =&  \widetilde{\tau} + \sum_{j\neq 0}\widetilde{\beta}_j (\mathbf{1}[\Delta D_{ij}=1] - \mathbf{1}[\Delta D_{ij} = -1]) \\
&+ \sum_{k}\sum_{j\neq 0, k}\widetilde{\delta}_{jk}\mathbf{1}[\Delta D_{ik}=1]\mathbf{1}[\Delta D_{ij}=-1] + \Delta \widetilde{\epsilon}_i,\nonumber\\
=& \widetilde{\tau} + \sum_{j\neq 0}\widetilde{\beta}_jD_{i00}D_{ij1}+\sum_{j\neq 0}(\widetilde{\delta}_{j0}-\widetilde{\beta}_{j})D_{ij0}D_{i01}\\
&+ \sum_{k}\sum_{j\neq 0, k}(\widetilde{\delta}_{jk}+\widetilde{\beta}_k-\widetilde{\beta}_j)D_{ij0}D_{ik1} + \Delta \widetilde{\epsilon}_i.\nonumber\label{eq:prop3sat}
 \end{align}
This is a saturated model for $E[\Delta Y_i \mid \{D_{ij0}D_{ik1}\}_{k\neq j}]$, with
\begin{align}
\widetilde{\beta}_j &= E[\Delta Y_i\mid D_{i00}D_{ij1}=1]-E[\Delta Y_i\mid D_{i\ell 0}D_{i\ell 1}=1, \forall \ell ],\\ 
\widetilde{\delta}_{j0}-\widetilde{\beta}_j & =E[\Delta Y_i\mid D_{ij0}D_{i01}=1]-E[\Delta Y_i\mid D_{i\ell 0}D_{i\ell 1}=1, \forall \ell ],\\
\widetilde{\delta}_{jk}+\widetilde{\beta}_k-\widetilde{\beta}_j & =E[\Delta Y_i\mid D_{ij0}D_{ik1}=1]-E[\Delta Y_i\mid D_{i\ell 0}D_{i\ell 1}=1, \forall \ell ].
\end{align}
Under Assumption IO and the parallel trends condition, we can write
\begin{align}
\widetilde{\beta}_j =& (E[Y_{i1}^j-Y_{i0}^0\mid D_{i00}D_{ij1}=1]-E[Y^0_{i1}-Y^0_{i0}\mid D_{i00}D_{i01}=1])p_0\nonumber \\
&+ (E[Y_{i1}^j-Y_{i0}^0\mid D_{i00}D_{ij1}=1]-E[Y^j_{i1}-Y^j_{i0}\mid D_{ij0}D_{ij1}=1])p_j\nonumber \\
&+\sum_{\ell\neq 0, j}(E[Y_{i1}^j-Y_{i0}^0\mid D_{i00}D_{ij1}=1]-E[Y^\ell_{i1}-Y^\ell_{i0}\mid D_{i\ell 0}D_{i\ell 1}=1])p_\ell\nonumber \\
=& E[Y_{i1}^j-Y_{i1}^0\mid D_{i00}D_{ij1}=1]p_0+E[Y_{i0}^j-Y_{i0}^0\mid D_{i00}D_{ij1}=1](1-p_0) \label{eq:prop3beta} \\ 
&+\sum_{\ell\neq 0, j}(E[Y_{i1}^j-Y_{i0}^j\mid D_{ij0}D_{ij1}=1]-E[Y^\ell_{i1}-Y^\ell_{i0}\mid D_{i\ell 0}D_{i\ell 1}=1])p_\ell,\nonumber
\end{align}
where $p_k=P(D_{ik0}=1\mid D_{i\ell0}=D_{i\ell1},\forall \ell)$. Similarly, we have
\begin{align}
\widetilde{\delta}_{j0}=& \widetilde{\beta}_j-(E[Y_{i1}^j-Y_{i1}^0\mid D_{ij0}D_{i01}=1]p_0+E[Y_{i0}^j-Y_{i0}^0\mid D_{ij0}D_{i01}=1](1-p_0))\nonumber \\
&-\sum_{\ell\neq 0, j}(E[Y_{i1}^j-Y_{i0}^j\mid D_{ij0}D_{ij1}=1]-E[Y^\ell_{i1}-Y^\ell_{i0}\mid D_{i\ell 0}D_{i\ell 1}=1])p_\ell\nonumber \\
=&E[Y_{i1}^{j}-Y_{i1}^{0}\mid D_{i00}D_{ij1}=1]p_{0}+E[Y_{i0}^{j}-Y_{i0}^{0}\mid D_{i00}D_{ij1}=1](1-p_{0})\\
&-\left(E[Y_{i1}^{j}-Y_{i1}^{0}\mid D_{ij0}D_{i01}=1]p_{0}+E[Y_{i0}^{j}-Y_{i0}^{0}\mid D_{ij0}D_{i01}=1])(1-p_{0})\right)\nonumber
\end{align}
and
\begin{align}
\widetilde{\delta}_{jk}=&\widetilde{\beta}_j-\widetilde{\beta}_k+E[Y_{i1}^k-Y_{i1}^j\mid D_{ij0}D_{ik1}=1]p_j + E[Y_{i0}^k-Y_{i0}^j\mid D_{ij0}D_{ik1}=1](1-p_j)\nonumber \\
&+\sum_{\ell\neq j, k}(E[Y_{i1}^k - Y_{i0}^k \mid D_{i\ell 0}D_{i\ell 1}=1, \forall \ell ])- E[Y_{i1}^\ell - Y_{i0}^\ell \mid D_{i\ell 0}D_{i\ell 1}=1, \forall \ell ])p_\ell\nonumber \\
=& E[Y_{i1}^j-Y_{i1}^0\mid D_{i00}D_{ij1}=1]p_0+E[Y_{i0}^j-Y_{i0}^0\mid D_{i00}D_{ij1}=1](1-p_0) \nonumber \\
&+\sum_{\ell\neq 0, j}(E[Y_{i1}^j-Y_{i0}^j\mid D_{ij0}D_{ij1}=1]-E[Y^\ell_{i1}-Y^\ell_{i0}\mid D_{i\ell 0}D_{i\ell 1}=1])p_\ell\nonumber\\
&- E[Y_{i1}^k-Y_{i1}^0\mid D_{i00}D_{ik1}=1]p_0-E[Y_{i0}^k-Y_{i0}^0\mid D_{i00}D_{ik1}=1](1-p_0) \nonumber \\
&-\sum_{\ell\neq 0, k}(E[Y_{i1}^k-Y_{i0}^k\mid D_{ik0}D_{ik1}=1]-E[Y^\ell_{i1}-Y^\ell_{i0}\mid D_{i\ell 0}D_{i\ell 1}=1])p_\ell,\nonumber\\
&+E[Y_{i1}^k-Y_{i1}^j\mid D_{ij0}D_{ik1}=1]p_j + E[Y_{i0}^k-Y_{i0}^j\mid D_{ij0}D_{ik1}=1](1-p_j)\nonumber \\
&+\sum_{\ell\neq j, k}(E[Y_{i1}^k - Y_{i0}^k \mid D_{i\ell 0}D_{i\ell 1}=1, \forall \ell ])- E[Y_{i1}^\ell - Y_{i0}^\ell \mid D_{i\ell 0}D_{i\ell 1}=1, \forall \ell ])p_\ell\nonumber \\
=& (E[Y_{i1}^j-Y_{i1}^0\mid D_{i00}D_{ij1}=1]+E[Y_{i0}^k-Y_{i0}^j\mid D_{ij0}D_{ik1}=1])p_0 \label{eq:prop3deltak} \\ 
&- E[Y_{i1}^k-Y_{i1}^0\mid D_{i00}D_{ik1}=1]p_0 \nonumber \\
&+(E[Y_{i0}^j-Y_{i0}^0\mid D_{i00}D_{ij1}=1]+E[Y_{i1}^k-Y_{i1}^j\mid D_{ij0}D_{ik1}=1])p_j\nonumber\\
&-E[Y_{i0}^k-Y_{i0}^0\mid D_{i00}D_{ik1}=1])p_j\nonumber \\
&+(E[Y_{i0}^j-Y_{i0}^0\mid D_{i00}D_{ij1}=1]+ E[Y_{i0}^k-Y_{i0}^j\mid D_{ij0}D_{ik1}=1])(1-p_0-p_j)\nonumber \\
& -E[Y_{i0}^k-Y_{i0}^0\mid D_{i00}D_{ik1}=1](1-p_0-p_j) \nonumber \\
&+\sum_{\ell\neq 0, j}(E[Y_{i1}^j-Y_{i0}^j\mid D_{ij0}D_{ij1}=1]-E[Y^\ell_{i1}-Y^\ell_{i0}\mid D_{i\ell 0}D_{i\ell 1}=1])p_\ell\nonumber\\
&-\sum_{\ell\neq 0, k}(E[Y_{i1}^k-Y_{i0}^k\mid D_{ik0}D_{ik1}=1]-E[Y^\ell_{i1}-Y^\ell_{i0}\mid D_{i\ell 0}D_{i\ell 1}=1])p_\ell,\nonumber\\
&+\sum_{\ell\neq j, k}(E[Y_{i1}^k - Y_{i0}^k \mid D_{i\ell 0}D_{i\ell 1}=1, \forall \ell ])- E[Y_{i1}^\ell - Y_{i0}^\ell \mid D_{i\ell 0}D_{i\ell 1}, \forall \ell ])p_\ell\nonumber
\end{align}
Finally, note that we can the standard omitted-variables bias formula to write the vector of mover regression coefficients in terms of the saturated model's coefficient vector:
\begin{align}
\beta &= \widetilde{\beta} + \sum_k\sum_{j\neq 0,k}\widetilde{\delta}_{jk}R_{jk},
\end{align}
where $R_{jk}$ denotes the coefficient vector from regressing each $\mathbf{1}[\Delta D_{ik} = 1]\mathbf{1}[\Delta D_{ij}=-1]$ on the set of $\Delta D_{i\ell}$ for $\ell>0$. Substituting equations (\ref{eq:prop3beta})-(\ref{eq:prop3deltak}) in to this expression shows that $\beta$ will not in general identify a weighted average of causal parameters.
$\hfill\square$

\pagebreak{}

\subsubsection*{Proof of Proposition 3 and Corollary}

Let $\psi_{i} = P(\Delta D_{i}\neq 0)/P(\Delta D_{i}\neq 0\mid X_i)$. For any two treatments $c$ and $d$ we have
\begin{align}
E[\Delta Y_i \rho_{it}^{c,d}\psi_{i}\mid X_i] &=
(-1)^t E\left[\Delta Y_i D_{ic,1-t}\frac{D_{ict}E[D_{idt}D_{ic,1-t}\mid X_{i}]-D_{idt}E[D_{ict}D_{ic,1-t}\mid X_{i}]}{E[D_{ict}D_{ic,1-t}\mid X_{i}]E[D_{idt}D_{ic,1-t}\mid X_{i}]}\mid X_i\right]\nonumber \\
&=(-1)^t \left(E\left[\Delta Y_i \mid D_{ic,1-t}D_{ict}=1,X_i\right] - E\left[\Delta Y_i \mid D_{ic,1-t}D_{idt}=1,X_i\right]\right)\nonumber \\
&=E[Y_{i,1-t}^c-Y_{it}^c \mid D_{ic,1-t}D_{ict}=1,X_i] - E[Y_{i,1-t}^c-Y_{it}^d \mid D_{ic,1-t}D_{idt}=1,X_i]\nonumber \\
&=E[Y_{it}^d-Y_{it}^c \mid  D_{ic,1-t}D_{idt}=1,X_i]\nonumber\\
&=E[Y_{it}^d-Y_{it}^c \mid  \Delta D_i \neq 0,X_i],\label{eq:prop3pf1}
\end{align}
provided $E[D_{ict}D_{ic,1-t}\mid X_{i}]E[D_{idt}D_{ic,1-t}\mid X_{i}]\neq 0$. Here the second-to-last line follows from Assumption CPT and, for $t=0$, Assumption COI, while the last line follows by Assumption CEH. The same steps and assumptions show that 
\begin{align}
E[\Delta Y_i (-\rho_{it}^{d,c})\psi_{i}\mid X_i]&=E[Y_{it}^d-Y_{it}^c \mid  D_{id,1-t}D_{ict}=1,X_i]\nonumber \\
&=E[Y_{it}^d-Y_{it}^c \mid  \Delta D_i \neq 0,X_i].\label{eq:prop3pf2}
\end{align}
Given the chain $C_\ell$ and set of constants $w_m$, we therefore have the main result. 
\begin{align}
MATE_{jt}  &= \int E[Y_{it}^j-Y_{it}^0 \mid \Delta D_{i}\neq0,  X_i] dP(X_i\mid\Delta D_{i}\neq0)\nonumber \\
&=\int  \left(\sum_{m=1}^{M_{j\ell}}  E[Y_{it}^{c_{j\ell m}}-Y_{it}^{c_{j\ell m-1}} \mid \Delta D_{i}\neq0,  X_i]\right) dP(X_i\mid\Delta D_{i}\neq0)\nonumber\\
&= E\left[\left(\sum_{m=1}^{M_{j\ell}} (w_m\Delta Y_i\rho_{it}^{c_{j\ell,m-1},c_{j\ell m}}\psi_i+(1-w_m)\Delta Y_i(-\rho_{it}^{c_{j\ell m},c_{j\ell,m-1}})\psi_i)\right)\frac{P(\Delta D_{i}\neq 0\mid X_i)}{P(\Delta D_{i}\neq0)}\right]\nonumber\\
&= E\left[\Delta Y_i \left(\sum_{m=1}^{M_{j\ell}}(w_{m}\rho_{it}^{c_{j\ell,m-1},c_{j\ell m}}+(1-w_{m})(-\rho_{it}^{c_{j\ell m},c_{j\ell,m-1}}))\right)\right].
\end{align}
The $J=2$ corollary follows from the second-to-last lines of equations (\ref{eq:prop3pf1}) and (\ref{eq:prop3pf2}) by noting that
\begin{align}
MATE_{1t}  =& \int E[Y_{it}^1-Y_{it}^0 \mid \Delta D_{i1}\neq0,  X_i] dP(X_i\mid\Delta D_{i1}\neq0)\nonumber \\
= & \int E[\Delta Y_i \rho_{it}^{0,1}\psi_{i}\mid X_i]\frac{E[D_{i1t}D_{i0,1-t}\mid X_i]}{P(\Delta D_{i1}\neq 0 \mid X_i)}dP(X_i\mid\Delta D_{i1}\neq0)\nonumber\\
&\quad +\int E[\Delta Y_i (-\rho_{it}^{1,0})\psi_{i}\mid X_i]\frac{E[D_{i0t}D_{i1,1-t}\mid X_i]}{P(\Delta D_{i1}\neq 0 \mid X_i)}dP(X_i\mid\Delta D_{i1}\neq0)\nonumber\\
= & E\left[\Delta Y_i \rho_{it}^{0,1}\psi_{i}\frac{E[D_{i1t}D_{i0,1-t}\mid X_i]}{P(\Delta D_{i1}\neq0)}
+\Delta Y_i (-\rho_{it}^{1,0})\omega_{i}\frac{E[D_{i0t}D_{i1,1-t}\mid X_i]}{P(\Delta D_{i1}\neq0)}\right]\nonumber\\
=& E\left[\Delta Y_i\left(w_{it}^*\rho_{it}^{0,1}+(1-w_{it}^*)\rho_{it}^{1,0}\right)\right].\quad\quad\quad\quad\quad\quad\quad\quad\quad\quad\quad\quad\quad\quad\quad\quad\quad\quad\square
\end{align}

\pagebreak{}

\subsubsection*{Proof of Proposition 4}
Let $\psi_{i} = P(\Delta D_{i}\neq 0)/P(\Delta D_i\neq 0 \mid X_i)$. For any two treatments $c$ and $d$ we have
\begin{align}
E[\Delta Y_i \kappa_{i}^{c,d}\psi_{i}\mid X_i] =&
E\left[\Delta Y_i \frac{1}{2}\frac{D_{ic0}D_{id1}E[D_{id0}D_{ic1}\mid X_{i}]-D_{id0}D_{ic1}E[D_{ic0}D_{id1}\mid X_{i}]}{E[D_{id0}D_{ic1}\mid X_{i}]E[D_{ic0}D_{id1}\mid X_{i}]}\mid X_i\right]\nonumber \\
=&\frac{1}{2}\left(E[\Delta Y_i \mid D_{ic0}D_{id1}=1,X_i] - E[\Delta Y_i \mid D_{id0}D_{ic1}=1,X_i]\right)\nonumber \\
=&\frac{1}{2}\left(E[Y_{i1}^d-Y_{i0}^c \mid D_{ic0}D_{id1}=1,X_i] - E[Y_{i1}^c-Y_{i0}^c \mid D_{ic0}D_{id1}=1,X_i]\right)\nonumber\\
&- \frac{1}{2}\left(E[Y_{i1}^c-Y_{i0}^d \mid D_{id0}D_{ic1}=1,X_i]-E[Y_{i1}^c-Y_{i0}^c \mid D_{id0}D_{ic1}=1,X_i]\right)\nonumber \\
=&\frac{1}{2}\left(E[Y_{i1}^d-Y_{i1}^c \mid D_{ic0}D_{id1}=1,X_i]+ E[Y_{i0}^d-Y_{i0}^c \mid D_{id0}D_{ic1}=1,X_i]\right)\nonumber\\
=&\frac{1}{2}\left(E[Y_{i1}^d-Y_{i1}^c \mid  \Delta D_i \neq 0,X_i]+E[Y_{i0}^d-Y_{i0}^c \mid  \Delta D_i \neq 0,X_i]\right),
\end{align}
provided $E[D_{id0}D_{ic1}\mid X_{i}]E[D_{ic0}D_{id1}\mid X_{i}]\neq 0$. Here the third equality follows from Assumptions CPT and COI, while the last line follows by Assumption CEH. Given the chain $C_\ell$ we thus have
\begin{align}
\frac{1}{2}(MATE_{j0}+MATE_{j1})  &= \int \frac{1}{2}\left(\sum_{t=0}^1 E[Y_{it}^j-Y_{it}^0 \mid \Delta D_{i}\neq0,  X_i] \right)dP(X_i\mid\Delta D_{i}\neq0)\nonumber \\
&=\int \frac{1}{2} \left(\sum_{t=0}^1 \sum_{m=1}^{M_{j\ell}}  E[Y_{it}^{c_{j\ell m}}-Y_{it}^{c_{j\ell m-1}} \mid \Delta D_{i}\neq0,  X_i]\right) dP(X_i\mid\Delta D_{i}\neq0)\nonumber\\
&= E\left[\left(\sum_{m=1}^{M_{j\ell}} \Delta Y_i\kappa_{i}^{c_{j\ell,m-1},c_{j\ell m}}\psi_i\right)\frac{P(\Delta D_{i}\neq 0\mid X_i)}{P(\Delta D_{i}\neq0)}\right]\nonumber\\
&= E\left[\Delta Y_i \left(\sum_{m=1}^{M_{j\ell}}\kappa_{i}^{c_{j\ell,m-1},c_{j\ell m}}\right)\right],
\end{align}
completing the proof. $\hfill$ $\square$

\pagebreak{}

\subsection*{Extension to Many Time Periods}

This appendix generalizes the MATE identification results to multi-period mover designs. Let $\Delta_{sr} V_i=V_{is}-V_{ir}$ denote the difference operator applied to variable $V_{it}$ over periods $s>r$, and let $\Delta D_i$ collect all $\Delta_{sr} D_{ij}$. The multi-period version of Assumptions CPT and COI are
\begin{description}
\item [{Assumption}] CPT$^\prime$: For each treatment $j$, periods $t>s$, and $x$ in the support of $X_i$, 
\begin{align}
E[Y_{it}^{\bar j_{t}\rightarrow j}-Y_{is}^{\bar j_{s}\rightarrow j}\mid D_{ijs}D_{ijt}=1,X_i=x]&=E[Y_{it}^{\bar j_{t}\rightarrow j}-Y_{is}^{\bar j_{s}\rightarrow j}\mid\Delta_{ts} D_{ij}=1,X_i=x]\\
&=E[Y_{it}^{\bar j_{t}\rightarrow j}-Y_{is}^{\bar j_{s}\rightarrow j}\mid\Delta_{ts} D_{ij}=-1,X_i=x],
\end{align}
where $\bar j_{r}$ denotes a vector of $j$'s of length $r$.
\item [{Assumption}] COI$^\prime$: For all treatments $(j,k)$, periods $t>s$, and $x$ in the support of $X_i$,
\begin{align}
E[Y_{it}^{\bar k_t\rightarrow j}\mid D_{iks}D_{ijt}=1,X_i=x]=&E[Y_{i1}^{\bar j_t \rightarrow j}\mid D_{iks}D_{ijt}=1,X_i=x],
\end{align}
where $\bar k_{r}$ denotes a vector of $k$'s of length $r$.
\end{description}
Here Assumption CPT$^\prime$ states that movers into or out of each treatment $j$ between periods $s$ and $t$ would have, had they stayed at $j$, followed the same average outcome trends as each other and as treatment $j$ stayers, conditional on the controls. Similarly Assumption COI$^\prime$ states that movers into each treatment $j$ from treatment $k$ between periods $s$ and $t$ have the same time-$t$ outcomes as if they had stayed at $j$. Assumption CEH from the text remains unmodified in the multi-period case. 

We then have the following result, the proof of which follows by the same steps as that of Proposition 3:
\begin{description}
\item [{Proposition}] 3$^\prime$: Suppose Assumptions CPT$^\prime$ and CEH hold, and for treatment $j$ and periods $t>s$ there exists a chain $C_{j\ell}$ such that, for each $m=1,\dots,M_{j\ell}$ and all $x$ in the support of $X_i$ either (i) $P(D_{i,m-1,s}D_{imt}=1\mid X_i=x)>0$ and $P(D_{i,m-1,s}D_{i,m-1,t}=1\mid X_i=x)>0$ or (ii) $P(D_{im,s}D_{i,m-1,t}=1\mid X_i=x)>0$ and $P(D_{ims}D_{imt}=1\mid X_i=x)>0$. Then
\begin{align}
MATE_{jt} &= E\left[\Delta_{ts} Y_i \left(\sum_{m=1}^{M_{j\ell}}(w_{m}\rho_{ist}^{c_{j\ell,m-1},c_{j\ell m}}+(1-w_{m})(-\rho_{ist}^{c_{j\ell m},c_{j\ell,m-1}}))\right)\right],\label{eq:tmate0}
\end{align}
where the $w_m$ are constants such that $w_m = 0$ if (i) fails for $m$ and $w_m = 1$ if (ii) fails for $m$, and where
\begin{align}
\tilde{\rho}_{it}^{c,d}=(-1)^{\mathbf{1}[t>s]} D_{ic,s}\frac{D_{ict}E[D_{idt}D_{ics}\mid X_{i}]-D_{idt}E[D_{ict}D_{ics}\mid X_{i}]}{E[D_{ict}D_{ics}\mid X_{i}]E[D_{idt}D_{ics}\mid X_{i}]}\frac{P(\Delta D_i\neq 0 \mid X_i)}{P(\Delta D_i\neq 0)}.\label{eq:tmate0_end}
\end{align} 
If moreover Assumption COI holds, equations (\ref{eq:mate0}) and (\ref{eq:mate0_end}) also hold for periods $t<s$.
\end{description}

We also have the following generalization of Proposition 4, the proof of which follows similarly
\begin{description}
\item [{Proposition}] 4$^\prime$: Suppose Assumptions CPT$^\prime$, CEH, and COI$^\prime$ hold, and for treatment $j$ and periods $(t,s)$ there exists a chain $C_{j\ell}$ such that, for each $m=1,\dots,M_{j\ell}$ and all $x$ in the support of $X_i$, $P(D_{i,m-1,s}D_{imt}=1\mid X_i=x)>0$ and $P(D_{ims}D_{i,m-1,t}=1\mid X_i=x)>0$. Then
\begin{align}
\frac{1}{2}(MATE_{js}+MATE_{jt}) &= E\left[\Delta Y_i \left(\sum_{m=1}^{M_{j\ell}}\kappa_{i}^{c_{j\ell,m-1},c_{j\ell m}}\right)\right],\label{eq:tcmate0}
\end{align}
where
\begin{align}
\tilde{\kappa}_{i}^{c,d}=\frac{1}{2}\frac{D_{ics}D_{idt}E[D_{ids}D_{ict}\mid X_{i}]-D_{ids}D_{ict}E[D_{ics}D_{idt}\mid X_{i}]}{E[D_{ids}D_{ict}\mid X_{i}]E[D_{ics}D_{idt}\mid X_{i}]}\frac{P(\Delta D_i\neq 0 \mid X_i)}{P(\Delta D_i\neq 0)}.\label{eq:tcmate0_end}
\end{align} 
\end{description}

Analogous results for the specification tests and efficient estimators derived in Section 3.3 would again apply here, with overidentification resulting from either many chains or many time period pairs satisfying Propositions 3$^\prime$ and 4$^\prime$.

\pagebreak{}





\end{document}